\documentclass[twocolumn]{aastex631}
\usepackage{enumitem}
\usepackage{booktabs}
\usepackage[fleqn]{amsmath}
\usepackage{amssymb}

\received{2/28/2023} \revised{7/10/2023} \accepted{7/17/2023}
\shorttitle{Gamma-Ray Pulsars in View of 3D Kinetic Global Magnetosphere Models}
\shortauthors{Kalapotharakos et al.} 

\newcommand{\g}{\gamma}

\newcommand{\ec}{\epsilon_{\rm cut}}

\newcommand{\ed}{\dot{\mathcal{E}}}
\newcommand{\eacc}{E_{\rm acc}}

\newcommand{\ir}{\mathcal{F}}

\begin{document}
\tighten

\title{The Gamma-Ray Pulsar Phenomenology in View of 3D Kinetic Global Magnetosphere Models}

\correspondingauthor{Constantinos Kalapotharakos}
\email{constantinos.kalapotharakos@nasa.gov,\\ ckalapotharakos@gmail.com}

\author[0000-0003-1080-5286]{Constantinos Kalapotharakos}
\affiliation{University of Maryland, College Park (UMCP/CRESST II)\\
College Park, MD 20742, USA} \affiliation{Astrophysics Science
Division, NASA/Goddard Space Flight Center\\ Greenbelt, MD 20771,
USA}

\author[0000-0002-9249-0515]{Zorawar Wadiasingh}
\affiliation{University of Maryland, College Park (UMCP/CRESST II)\\
College Park, MD 20742, USA} \affiliation{Astrophysics Science
Division, NASA/Goddard Space Flight Center\\ Greenbelt, MD 20771,
USA}

\author[0000-0001-6119-859X]{Alice K. Harding}
\affiliation{Theoretical Division, Los Alamos National Laboratory\\ Los Alamos, NM 87545, USA}

\author[0000-0002-7435-7809]{Demosthenes Kazanas} \affiliation{Astrophysics Science
Division, NASA/Goddard Space
Flight Center\\ Greenbelt, MD 20771, USA} 




\begin{abstract}
{We develop kinetic plasma models of pulsar magnetospheres with magnetic-field-line-dependent plasma injection that reveal the importance of various magnetosphere regions in regulating the $\gamma$-ray emission. We set different particle injection rates for the so-called open, closed, and separatrix zones. Moderate particle injection rates in open and closed zones ensure a global field structure close to the force-free one, while the dissipation occurs mainly in and around the equatorial current sheet. The particles injected in the separatrix zone affect the particle populations that enter the equatorial current sheet region and, therefore, the corresponding accelerating electric fields, particle energies, the spectral cutoff energy, and $\gamma$-ray efficiency. The separatrix zone models reproduce the recently discovered fundamental plane of $\gamma$-ray pulsars consistent with curvature radiation emission, the $\gamma$-ray light-curve shapes, and the radio-lag vs. peak-separation correlation reported in the Fermi second pulsar catalog. The model beaming factors indicate that the pulsar total $\gamma$-ray luminosities listed in the Fermi catalogs are overestimations of the actual ones. We find that the radiation reaction limited regime starts ceasing to govern the high-energy emission for $\ed\lesssim 10^{34}\rm \, \, erg\,\;s^{-1}$. Our results also indicate that toward high magnetic inclination angles, the ``Y point" around the rotational equator migrates well inside the light cylinder sparking additional peaks in the $\gamma$-ray pulse profiles. We find that an equivalent enhanced particle injection beyond the Y point strengthens these features making the model $\gamma$-ray light curves inconsistent with those observed.}
\end{abstract}



\defcitealias{2013ApJS..208...17A}{2PC}
\defcitealias{2020ApJS..247...33A}{4FGL}
\defcitealias{2022arXiv220111184F}{4FGL-DR3}
\defcitealias{2005AJ....129.1993M}{ATNF}
\defcitealias{2019ApJ...883L...4K}{K19}
\defcitealias{2018ApJ...857...44K}{K18}

\section{Introduction} \label{sec:intro}

Over the past decade, the Fermi Large Area Telescope (LAT) has greatly increased the number of $\gamma$-ray rotation-powered pulsars detected between 30 MeV–300 GeV. The LAT has detected over 270 new $\gamma$-ray pulsars to date [117 of which were compiled in the Second Fermi Pulsar Catalog \citep[2PC,][]{2013ApJS..208...17A} and more expected later this year in the forthcoming Third Pulsar Catalog (3PC).

The Fermi data largely ruled out the possibility that the observed $\gamma$-rays originate from near the stellar surface, implicating the outer magnetosphere as the region of the pulsar $\gamma$-ray emission in most pulsars.
Moreover, the patterns of the $\gamma$-ray light curves show clear dependence of the $\gamma$-ray pulse peak separation $\Delta$ on the
phase-lag between radio and $\gamma$-ray emission $\delta$. This important multiwavelength correlation sets strict constraints on the pulsar emission physics and geometry
\citep{2010MNRAS.404..767C,2010ApJ...715.1282B,2014ApJ...793...97K}. These studies indicated the equatorial current sheet (ECS), a characteristic feature of the force-free (FF) global magnetosphere solutions \citep{1999ApJ...511..351C,2006MNRAS.368.1055T,2006ApJ...648L..51S} emerging at and beyond the light-cylinder (LC), as the prime candidate region for the observed $\gamma$-ray pulsar emission.

\citet{2017ApJ...842...80K}, assuming curvature radiation in
the radiation reaction limited (RRL) regime that occurs at the ECS region near the LC, demonstrated that the
LAT data imply the corresponding accelerating
electric-field components, $\eacc$, and their dependence on the
spin-down power $\ed$, uncovering the corresponding operational regime, i.e., the dissipation degree of the emitting region.

More recently, \citet{2019ApJ...883L...4K}, using 88 $\gamma$-ray pulsars from the \citetalias{2013ApJS..208...17A}, showed that the entire
$\gamma$-ray pulsar population, i.e., millisecond pulsars (MPs) and
young pulsars (YPs), lie on a fundamental plane (FP)
\begin{equation}
\label{eq:fp_observations} L_{\gamma}\propto \ec^{1.18\pm
0.24}B_{\star}^{0.17\pm 0.05}\ed^{0.41\pm 0.08}
\end{equation}
that relates their total $\gamma$-ray luminosity, $L_{\gamma}$,
spectral cutoff energy, $\ec$, stellar surface magnetic field,
$B_{\star}$, and spin-down power, $\ed$. Remarkably, the observed FP
is consistent with the theoretically expected behavior
\begin{equation}
\label{eq:fp_theory} L_{\gamma}\propto
\ec^{4/3}B_{\star}^{1/6}\ed^{5/12}
\end{equation}
that attributes the pulsar $\gamma$-ray emission to curvature radiation in the RRL\footnote{\citet{2022ApJ...934...65K} showed that the RRL regime is a sufficient but not a necessary condition for the derivation of the FP theoretical scaling, i.e., Eq.~(\ref{eq:fp_theory}).} by particles accelerated in the ECS just outside the LC.

\begin{figure}[]
\vspace{0.0in}
  \begin{center}
    \includegraphics[width=0.4\textwidth]{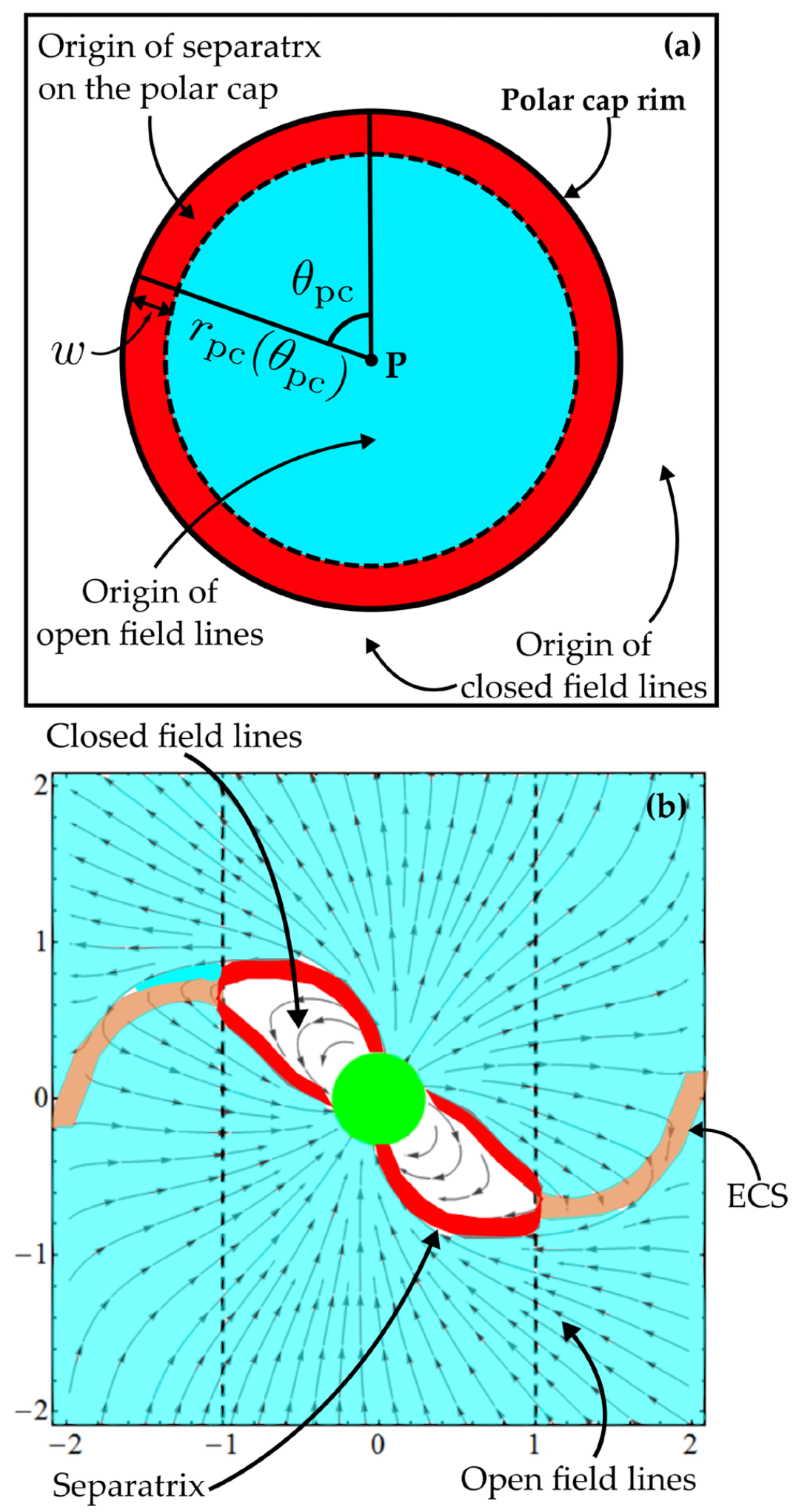}
  \end{center}
  \vspace{-0.2in}
  \caption{The \textit{separatrix zone} model is schematically presented. \textbf{Top panel:} The polar-cap region around a magnetic pole is denoted by the point P. The polar-cap rim is the origin of the last open magnetic field lines, i.e., those that do not cross the LC. The separatrix zone is the region along all the magnetic field lines that originate at the red zone. The width, $w$, of the red zone is measured as a fraction of the polar cap radius, $r_{\rm pc}$. In simulations, the polar cap rim, i.e., $r_{\rm pc}(\theta_{\rm pc})$, is identified (at every time step) as the origin of the field lines that reach the LC. \textbf{Bottom panel:} The FF magnetosphere structure on the poloidal magnetic dipole moment $-$ angular velocity, ($\pmb{\mu}-\pmb{\Omega}$), plane. The red region denotes the separatrix zone, while the blue and white regions denote the zones of the open and closed magnetic field lines, respectively. The orange region that emanates from the tip of the separatrix zone denotes the ECS region.}
  \label{fig:001}
  \vspace{-0.0in}
\end{figure}

\citet{2022ApJ...934...65K}\footnote{They adopted the spectral analysis data from the 12-year incremental version of the fourth Fermi-LAT catalog of point $\gamma$-ray sources (4FGL, \citealt{2020ApJS..247...33A}; 4FGL-DR3, for Data Release 3, \citealt{2022arXiv220111184F}) combined with data from the Australia Telescope National Facility (ATNF) Pulsar Catalog
\citep{2005AJ....129.1993M}.} expanded the pulsar sample by more than two-fold to 190, which they used to update the fundamental plane relation, i.e., the pulsar $\gamma$-ray luminosity
\begin{equation}
\label{eq:fp_4FGL} L_{\gamma}=10^{14.3\pm 1.3}\epsilon_{\rm cut}^{1.39\pm
0.17}B_{\star}^{0.12\pm 0.03}\ed^{0.39\pm 0.05}
\end{equation}
where $\epsilon_{\rm cut}$ is measured in MeV, $B_{\star}$ is measured in G, and $\ed$ and $L_{\gamma}$ are measured in $\rm erg\;s^{-1}$. The updated FP relation in Eq.~\eqref{eq:fp_4FGL} is compatible with both the FP relation derived using the \citetalias{2013ApJS..208...17A} data, i.e., Eq.~\eqref{eq:fp_observations} and the theoretically predicted behavior for curvature radiation, i.e., Eq.~\eqref{eq:fp_theory}.

This luminosity expression has also been independently corroborated by
\citet{2020JCAP...12..035P} who studied MPs exploring the possibility that unresolved MPs are responsible for the observed galactic center excess (\citealt{Abazajian2011}; also see, \citealt{2015ApJ...812...15B,2016PhRvL.116e1102B,Hooper_2016,2017ApJ...840...43A,2018ApJ...863..199G,2021arXiv210908439P,2021PhRvD.104d3007B,2021arXiv210600222G}).

\begin{figure}
\vspace{0.0in}
  \begin{center}
    \includegraphics[width=0.45\textwidth]{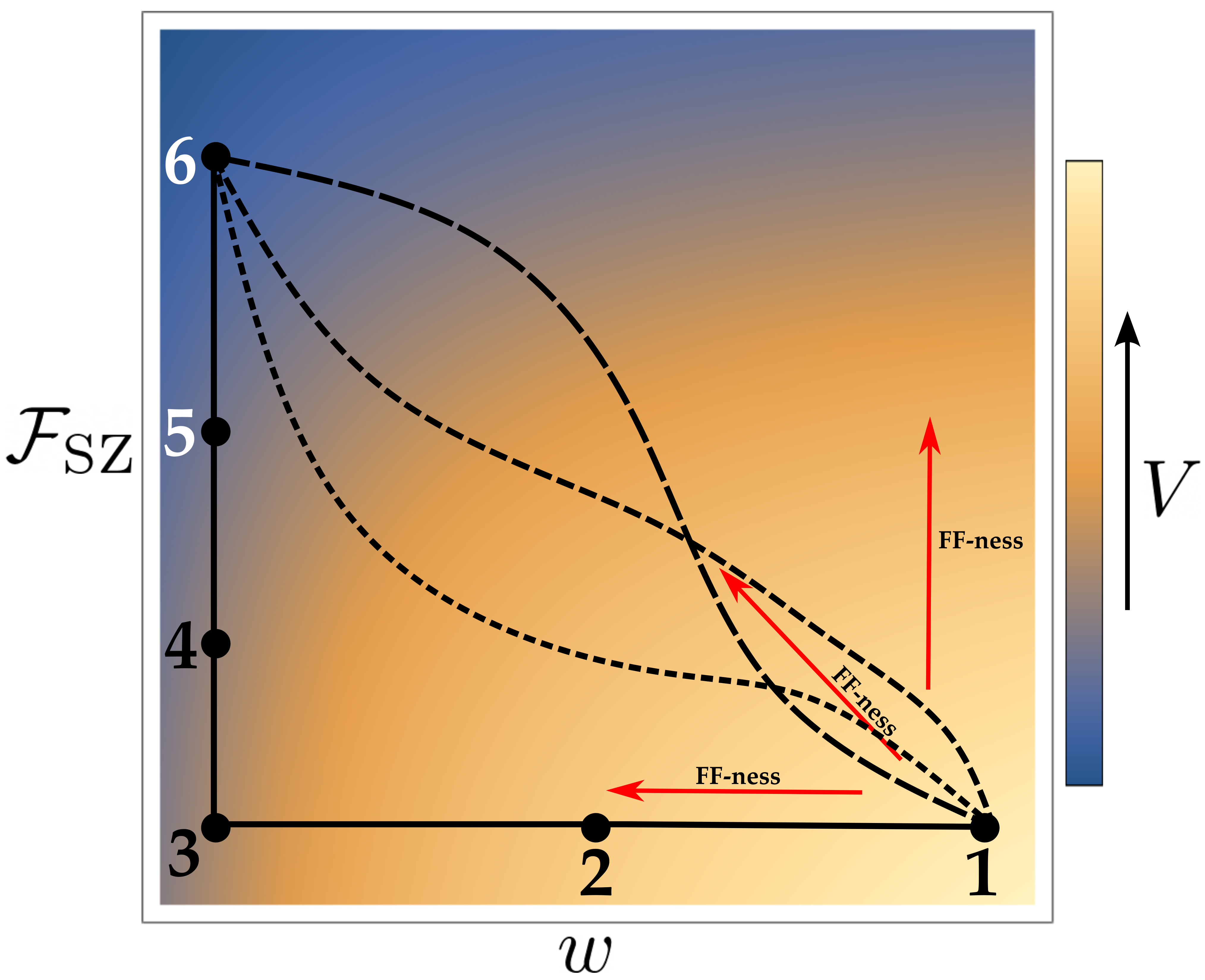}
  \end{center}
  \vspace{-0.2in}
  \caption{A schematic representation of the voltage, $V$, the particles encounter, in the indicated color scale, as a function of $w$ and $\mathcal{F}_{\rm SZ}$. The red arrows show the direction toward higher FF-ness. The big dots show our models on the $\ir_{\rm SZ}-w$ space, and each of them is numbered according to the numbers of our models (see Table~\ref{tab:sz_models}). We also note that the solid line that connects them shows the covered pathway schematically and is not related to the axes of the $(w, \ir_{\rm SZ})$ space. The various dashed lines demonstrate different pathways that trace the same FF-ness, i.e., $V$ values.}
  \label{fig:dissipative_character_schematic}
  \vspace{0.0in}
\end{figure}

\citet{2018ApJ...857...44K}, \citetalias{2018ApJ...857...44K} hereafter, and \citet{2018ApJ...858...81B} generated
particle-in-cell (PIC) models of 3D global pulsar magnetospheres and
investigated a range of global particle injection rates, $\ir$, normalized to the Goldreich-Julian rate\footnote{The particle flow from the polar caps; see also Table~\ref{tab:sz_models}.} \citep{1969ApJ...157..869G}, producing an entire spectrum of solutions from near vacuum retarded dipole up to near
force-free. In \citetalias{2018ApJ...857...44K}, the particle injection
occurred everywhere in the magnetosphere while in \citet{2018ApJ...858...81B}
only close to the stellar surface. Nonetheless, in both studies, the
corresponding particle injection prescriptions were agnostic to the magnetic field geometry, which is not believed to be true in real pulsars. Moreover, \citetalias{2018ApJ...857...44K}, by rescaling the particle energies appropriately to realistic values and demanding the $\ec$ of the corresponding $\gamma$-ray emission to match the observed values,
revealed a monotonic dependence of $\ir$ on $\ed$.

\begin{figure*}[]
\vspace{0.0in}
  \begin{center}
    \includegraphics[width=0.99\textwidth]{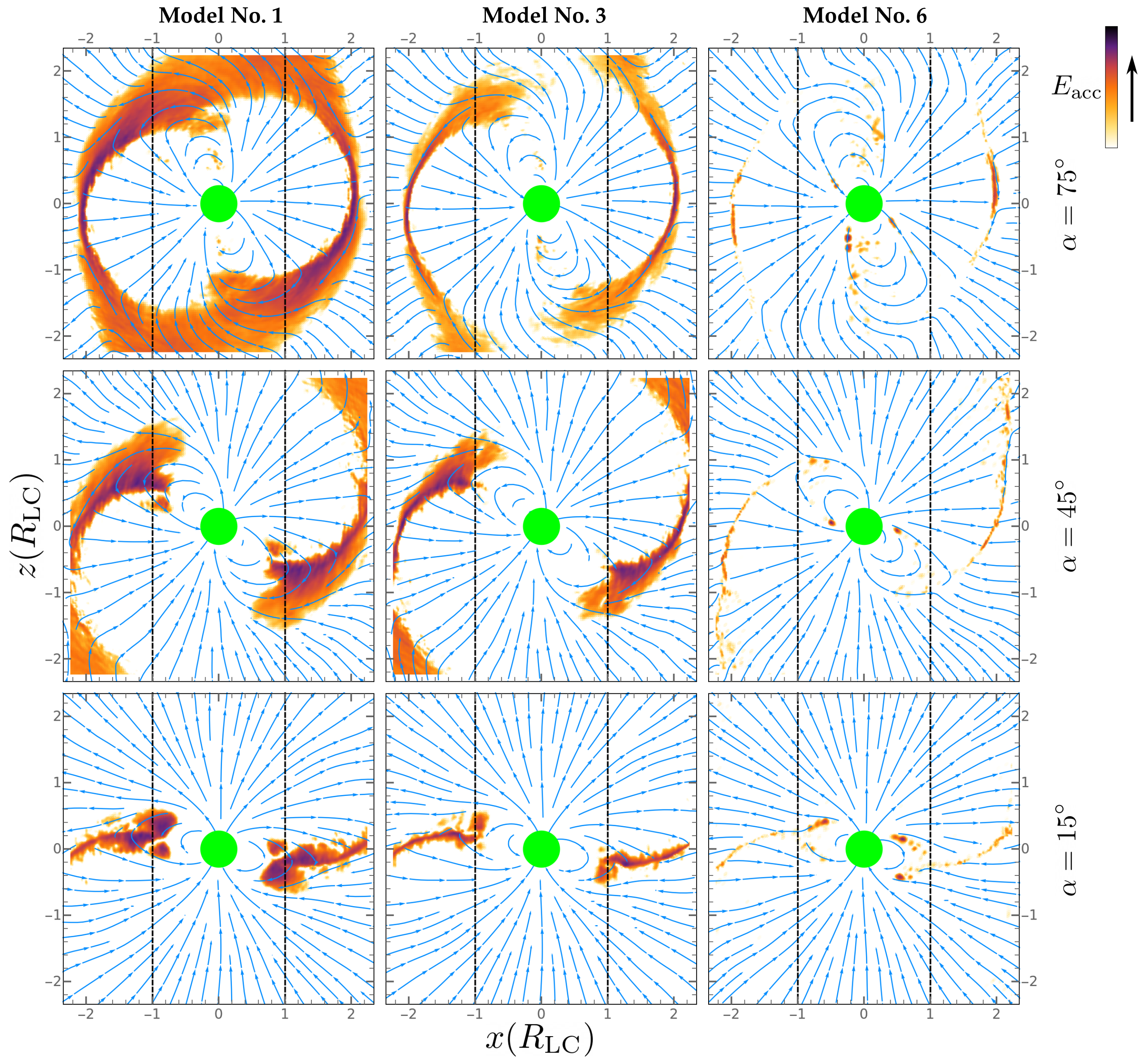}
  \end{center}
  \vspace{-0.3in}
  \caption{The high accelerating regions for the indicated $\alpha$ and model number values. The color scale denotes the accelerating electric field components.}
  \label{fig:002}
  \vspace{0.0in}
\end{figure*}

In spite of the success of the above models in reproducing the observed $\ec$,
for lower $\ir$ values--that are required for low $\ed$ values--additional features began appearing in the model sky maps of the $\gamma$-ray emission, and the corresponding $\gamma$-ray light curves were not always consistent
with the observed ones, e.g., Figure~17
in \citetalias{2018ApJ...857...44K}. This was due to the uniform decrease of $\ir$  (along all
the magnetic field flux tubes) that affected the global magnetospheric solution. The solutions corresponding to the lower $\ir$ values
started to deviate from the FF ones, introducing accelerating electric field components, $\eacc$, in wider and
different magnetosphere regions producing the additional emission
components highly inconsistent with the observed morphology of the $\gamma$-ray pulse profiles.

In reality, particle injection in pulsars at low altitudes occurs through single photon pair cascades on extremely small scales compared to the LC \citep{2013MNRAS.429...20T,2015ApJ...810..144T}. These are not resolvable from first principles in any global model (in a realistic manner), and thus only observations can inform on the underlying locales for PIC models. Moreover, the FF polar-cap regions are divided into sub-regions of different $J/\rho_{\rm GJ}$ values, where $J$ and $\rho_{\rm GJ}$ are the current density and the Goldreigh-Julian charge density, that, in principle, correspond to different pair-creation efficiencies \citep{2013MNRAS.429...20T}. The agnostic plasma injection schemes of \citet{2018ApJ...858...81B}, \citetalias{2018ApJ...857...44K} did not allow a detailed study of more physical field-dependent injection schemes that regulate particle acceleration and model $\gamma$-ray emission. Finally, these models were never tested against the FP because this had not been discovered yet.

\begin{figure}[!tbh]
\vspace{0.0in}
  \begin{center}
    \includegraphics[width=0.78\linewidth]{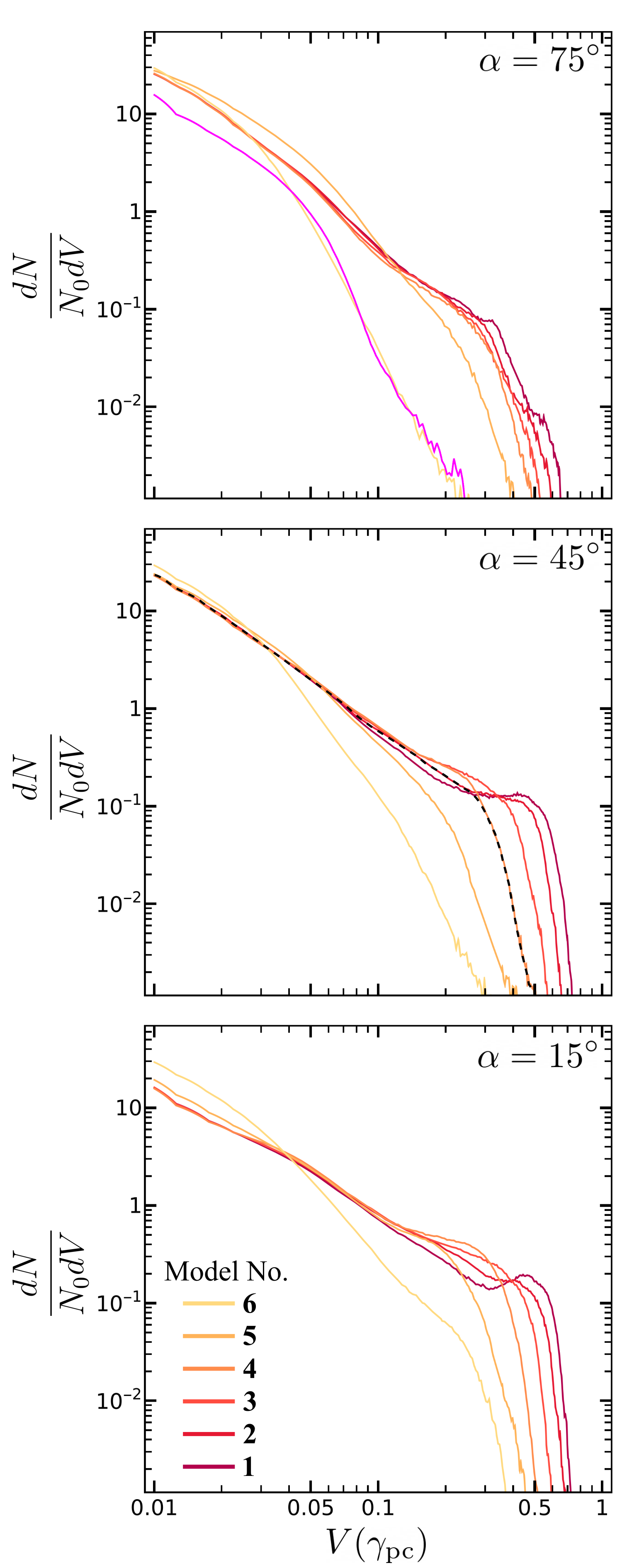}
  \end{center}
  \vspace{-0.2in}
  \caption{{The normalized particle distributions of the encountered voltage, where $dN$ denotes the number of particles within $V$ and $V+dV$ and $N_0$ denotes the total number of particles in the simulation.} The $V$ values are measured in the voltage across the polar cap. Each panel corresponds to the indicated $\alpha$ values, while the indicated colors denote the different model numbers. The more FF a model, the lower the cutoff of the V distributions. The black dashed line in the middle panel corresponds to a simulation of $w=0.12$ and
$\ir_{\rm SZ}=1.5\ir_{\rm GJ}^0$, which is equivalent to model No. 4, i.e.,  $w=0.04$, and $\ir_{\rm SZ}=1.0\ir_{\rm GJ}^0$. The magenta line in the top panel denotes a simulation equivalent to model No. 6 in which particle injection occurs in the high $\eacc$ region beyond the Y-point (see \S~\ref{sec:ecs}).}
  \label{fig:voltage_distribution}
  \vspace{0.0in}
\end{figure}

In this paper, we introduce a series of new 3D global
PIC models that implement magnetic field line-dependent particle injection for the first time.
This approach allows us to reproduce (at an unprecedented level) the Fermi $\gamma$-ray pulsar phenomenology, i.e., $\gamma$-ray pulse profiles, spectra, and the FP,
which facilitates an advanced interpretation of the observations.

The structure of the paper is as follows. In Section~\ref{sec:SImodel}, we describe the magnetic field line-dependent injection models focusing on the so-called \emph{separatrix zone model}. In Section~\ref{sec:results}, we present our results and finally in Section~\ref{sec:conclusions}, we present our
conclusions and discuss their implications and significance.

\section{The Separatrix Zone Model}
\label{sec:SImodel}

The analysis of the results of \citetalias{2018ApJ...857...44K} and their
comparison with the Fermi data (see \S\ref{sec:intro})
as well as the findings of \cite{2014ApJ...793...97K,2017ApJ...842...80K,2019ApJ...883L...4K,2022ApJ...934...65K}
indicate that \textbf{(a)} the global field structure should not
deviate considerably from FF (even for low $\ed$) and
\textbf{(b)} the dissipative region where most of the pulsar
$\gamma$-ray emission is produced should always remain near the ECS
mainly beyond the LC.

The particles that enter the dissipative zone near the ECS are
mainly injected in the broader separatrix zone that separates the open and the closed field zone. The current flow in the separatrix is connected with the current flow in the ECS; therefore, the two zones are strongly coupled.

In order to isolate the dissipation in a relatively narrow zone
proximate to the ECS, the particle injection in most of the magnetosphere should be adequate to sustain ideal FF conditions.
\begin{enumerate}[label=\Alph*.]
    \item The \textit{separatrix zone} is defined as the broader magnetosphere region along the last open magnetic field lines that originate
        within a specified angular fraction, $w$, measured in units of the corresponding polar cap radius, (reddish regions
        in Figure~\ref{fig:001}),
    \item the \textit{open zone} that is defined as the magnetosphere region
        along the rest of the open magnetic field lines (aqua colored
        regions in Figure~\ref{fig:001}),
    \item the \textit{closed zone} that is defined as the magnetosphere region
        along the closed, within the LC, magnetic field lines (white
        regions in Figure~\ref{fig:001}).
\end{enumerate}

The above description actually implies that there are three relevant zones where plasma injection can govern both the global field structure and particle acceleration for $\gamma$-ray emission.

\begin{deluxetable*}{ccccccc}[]
\tablecaption{The \textit{separatrix zone injection} models\label{tab:sz_models}}
\tablewidth{0pt} \tablehead{
\multicolumn{1}{c}{} & \multicolumn{2}{c}{$\alpha=15^{\circ}$} &
\multicolumn{2}{c}{$\alpha=45^{\circ}$} &
\multicolumn{2}{c}{$\alpha=75^{\circ}$}\\
\cmidrule(lr){2-3}\cmidrule(lr){4-5}\cmidrule(lr){6-7} \colhead{} &
\colhead{$\ir_{\rm OZ}=5.0$} & \colhead{$\ir_{\rm CZ}=5.0$} &
\colhead{$\ir_{\rm OZ}=5.0$} & \colhead{$\ir_{\rm CZ}=4.0$} &
\colhead{$\ir_{\rm OZ}=5.0$} & \colhead{$\ir_{\rm
CZ}=8.0$}\\} 
\startdata  & $\ir_{\rm SZ}$ & $w$ & $\ir_{\rm SZ}$ & $w$ &
$\ir_{\rm SZ}$ & $w$\\
Model No. & $(\ir_{\rm GJ}^0)$ & ($r_{\rm pc}$) & $(\ir_{\rm GJ}^0)$
& ($r_{\rm pc}$) &
$(\ir_{\rm GJ}^0)$ & ($r_{\rm pc}$)\\
\hline
1 & 0.0 & 0.15 & 0.0 & 0.15 & 0.0 & 0.15 \\
2 & 0.0 & 0.10 & 0.0 & 0.10 & 0.0 & 0.10 \\
3 & 0.0 & 0.04 & 0.0 & 0.04 & 0.0 & 0.04 \\
4 & 1.0 & 0.04 & 1.0 & 0.04 & 1.0 & 0.04 \\
5 & 3.0 & 0.04 & 3.0 & 0.04 & 3.0 & 0.04 \\
6 & 7.0 & 0.04 & 7.0 & 0.04 & 12.0 & 0.04 \\
\enddata
\tablecomments{All the particle injection rates, $\ir$, are measured in units of $\ir_{\rm GJ}^0$, which corresponds to the Goldreich-Julian particle flow $\rho_{\rm GJ}S_{\rm pc}c$. Here $\rho_{\rm GJ}$ and $S_{\rm pc}$ are the polar-cap charge density and the polar-cap area, respectively, of the aligned rotator, i.e., $\alpha=0^{\circ}$. The width of the separatrix zone, $w$, is measured in units of the corresponding polar cap radius, $r_{\rm pc}$.}
\end{deluxetable*}

Nonetheless, the regulation of $\eacc$ and the corresponding spectral $\ec$ is
achieved by the number of particles that eventually lie in the
dissipative zone independent of where these particles were injected. Hence, the particles that regulate the $\eacc$ and the high-energy emission can be, in principle, injected in the ECS region, i.e., produced in situ,
\citep[e.g.,][]{2018ApJ...855...94P,2019ApJ...877...53H}.

Thus, a fourth zone could be considered in the region close to the ECS beyond the LC, denoted by the orange color in the bottom panel of Figure~\ref{fig:001}b. We consider this zone as an extended
part of the separatrix zone and is discussed below (see Section~\ref{sec:ecs}).

We have produced six \textit{separatrix zone} models for three
magnetic inclination angles\footnote{{The magnetic inclination angle, $\alpha$, is defined as the angle between the stellar angular velocity $\pmb{\Omega}$ and the dipole magnetic moment $\pmb{\mu}$.}} (i.e., $\alpha=15^{\circ},~45^{\circ},~75^{\circ}$). For
each $\alpha$ value, we keep the $\ir$ values in the open zone, i.e., $\ir_{\rm
OZ}$, and in the closed zone, i.e., $\ir_{\rm CZ}$, fixed and we change the $w$ and
$\ir$ of the separatrix zone, i.e., $\ir_{\rm SZ}$. The particle injection is implemented
similarly to \citetalias{2018ApJ...857...44K}, but for all the zones, it takes place
up to a spherical distance equal to $0.7R_{\rm LC}$, i.e., $r_{\rm inj}\leq
0.7R_{\rm LC}$\footnote{{However, this rule is violated when we implement particle injection beyond the Y point in the extended separatrix zone discussed below in Section~\ref{sec:ecs}.}}.
More specifically, for each cell in the simulation, up to $r\leq 0.7R_{\rm LC}$, we inject one pair ($e^-, e^+$) at rest if the plasma magnetization exceeds locally the value $\Sigma_0 (r_s/r)^3$, where $r_s$ is the stellar radius and $\Sigma_0$ is a self-regulating constant that gradually adjusts, i.e., increases/decreases until the predefined particle injection rate is achieved. Thus, three $\Sigma_0$ values are considered corresponding to the three different injection zones. In each time step, we find the polar cap radius $r_{\rm pc}$ for each $\theta_{\rm pc}$ (see Figure~\ref{fig:001}). We use the bisectional method to calculate the $r_{\rm pc}$ values. For that, we integrate magnetic field lines outwards until they reach either the LC or the stellar surface. The initial $r_{\rm pc}$ values are always, i.e., for each $\theta_{\rm pc}$, those corresponding to the magnetic axis (guaranteed open line) and to the magnetic polar angle of $85^\circ$ (guaranteed closed line). The derivation of the polar cap combined with the preselected $w$ value delineates the regions on the stellar surface where the magnetic field lines corresponding to the three different injection zones originate. With the polar cap at hand, we integrate inwards the magnetic field lines originating  at the cells with $r_s<r<0.7R_{\rm LC}$ until they reach the stellar surface. The trace of the magnetic field line on the stellar surface allows the determination of the zone the cell belongs to.
The adopted  $\ir_{\rm OZ}$ and $\ir_{\rm CZ}$ values, shown
in Table~\ref{tab:sz_models}, ensure FF conditions, i.e., FF global field structure and negligible $\eacc$,
in the corresponding zones for all the models. A model's proximity to the
ideal FF conditions depends on $w$ and $\ir_{\rm SZ}$. An
appropriate measure of the FF-ness, i.e., the ideal character of the magnetosphere, would be the potential drop $V$ the
high-energy particles encounter, which
reflects the effective integrated accelerating electric field, i.e., $\eacc$\footnote{Another measure of the FF-ness would
be the fraction of the dissipated energy. However, this quantity may decrease
towards the vacuum retarded dipole models, which are dissipation-less. On the other hand, the
more FF a model is, the less the $V$ is.}. In
Figure~\ref{fig:dissipative_character_schematic}, we present the behavior of the FF-ness, i.e., potential drop $V$, of the magnetosphere
model on $\ir_{\rm SZ}$ and $w$, schematically. The potential drop $V$, the accelerating
particles encounter in the dissipative region, increases with $w$ and
decreases with $\ir_{\rm SZ}$. The red arrows in Figure~\ref{fig:dissipative_character_schematic} indicate the growing
directions of FF-ness, i.e., decreasing $V$ values. The model FF-ness is considered the same for equal $V$
values. The particle density in the ECS, i.e., the dissipative region, which
regulates $V$, depends on the prescription of the supplied pairs. However,
the particle density in the ECS cannot be a priori chosen in PIC
simulations because the ECS, like any other magnetosphere region, is
developed self-consistently. Thus, even when no particles are
injected into the separatrix zone, the ECS is still formed by drawing particles from
neighboring regions. For specific $\ir_{\rm OZ}$ and $\ir_{\rm CZ}$ values,
there are, in principle, different combinations of $w$ and $\ir_{\rm SZ}$
that lead to the same $V$. Therefore, the filling of a
certain range of $V$ values can be achieved along various paths in the
$(w,~\ir_{\rm SZ})$ space as demonstrated by the different dashed lines that connect points 1 and 6 in Figure~\ref{fig:dissipative_character_schematic}. In our
approach, we have chosen to change only one parameter ($w$ or $\ir_{\rm SZ}$)
in each step, and so we have followed a path similar to the one depicted by
the solid line in Figure~\ref{fig:dissipative_character_schematic}. The black
dots along this line and the corresponding numbers indicate the 6 different
models that are shown in Table~\ref{tab:sz_models}. For our models, we have adopted the maximum $w$ value to be $w_{\max}=0.15$.

We find that the global field structures in all the models are close to the FF
ones and even for the less FF models, i.e., model 1, the $\ed$ are always
greater than $0.7-0.8$ times the corresponding FF values. All the models develop
significant $\eacc$ mainly near the ECS structure that emanates from the tip of
the closed zone.

Our approach allows us to concentrate our efforts on the zone, i.e., the separatrix zone, that is mainly responsible for the regulation of
$\gamma$-ray emission. At the same time, the small particle
injection rates required in the open zone and closed zone compared to the rates required for uniform injection make our
simulations numerically less demanding since the total number of particles remains within feasible limits .\footnote{At a glance, the particle injection in the open zone and closed zone are comparable to that in the separatrix zone. However, the particles in the separatrix zone are injected within a much smaller volume than those that are injected in the open zone and closed zone.} In FF field structures, there are no currents along the magnetic field lines in the closed zone. Nonetheless, this does not forbid current-less particle motions along these lines. Moreover, particles near the border of the closed zone can always interact with the separatrix zone and escape toward it. This becomes easier in simulations where the gyroradius, in $R_{\rm LC}$ units, is considerably larger than in real pulsar magnetospheres. Both these effects trigger the need for a continuous particle injection in the closed zone.

For our simulations, we use the PIC code C-3PA
\citepalias{2018ApJ...857...44K} following an approach similar to the one
described in \citetalias{2018ApJ...857...44K}. Thus, the adopted
simulation stellar surface magnetic field is $B_{\rm s}=10^6\rm \, G$
and rotational period is $P_{\rm s}=0.1\rm \, s$. The spatial
resolution is $0.02 R_{\rm LC}$, the time step is $\Delta t_{\rm
s}=5\times 10^{-5}P_{\rm s}$, and the stellar radius is at $r_{\rm
s}=0.28R_{\rm LC}$. The main cubical computational domain has an edge length $\sim 9 R_{\rm LC}$ while the typical number of particles per cell at the LC is 5-10. The skin depth $d_{\rm s}=c/\omega_{\rm p}$, where $\omega_{\rm p}$ the plasma frequency, is resolved for all the particles with energy above a few percent of the energy corresponding to the entire potential drop at the polar cap even for the models with the highest $\ir$ values. This
implies that the particle population with lower energies could be artificially heated to energies higher than expected physically. However, this particle population does not contribute to the high-energy emission even though it contributes to the screening of $\eacc$.

\begin{figure*}[!tbh]
\vspace{0.0in}
  \begin{center}
    \includegraphics[width=1.0\linewidth]{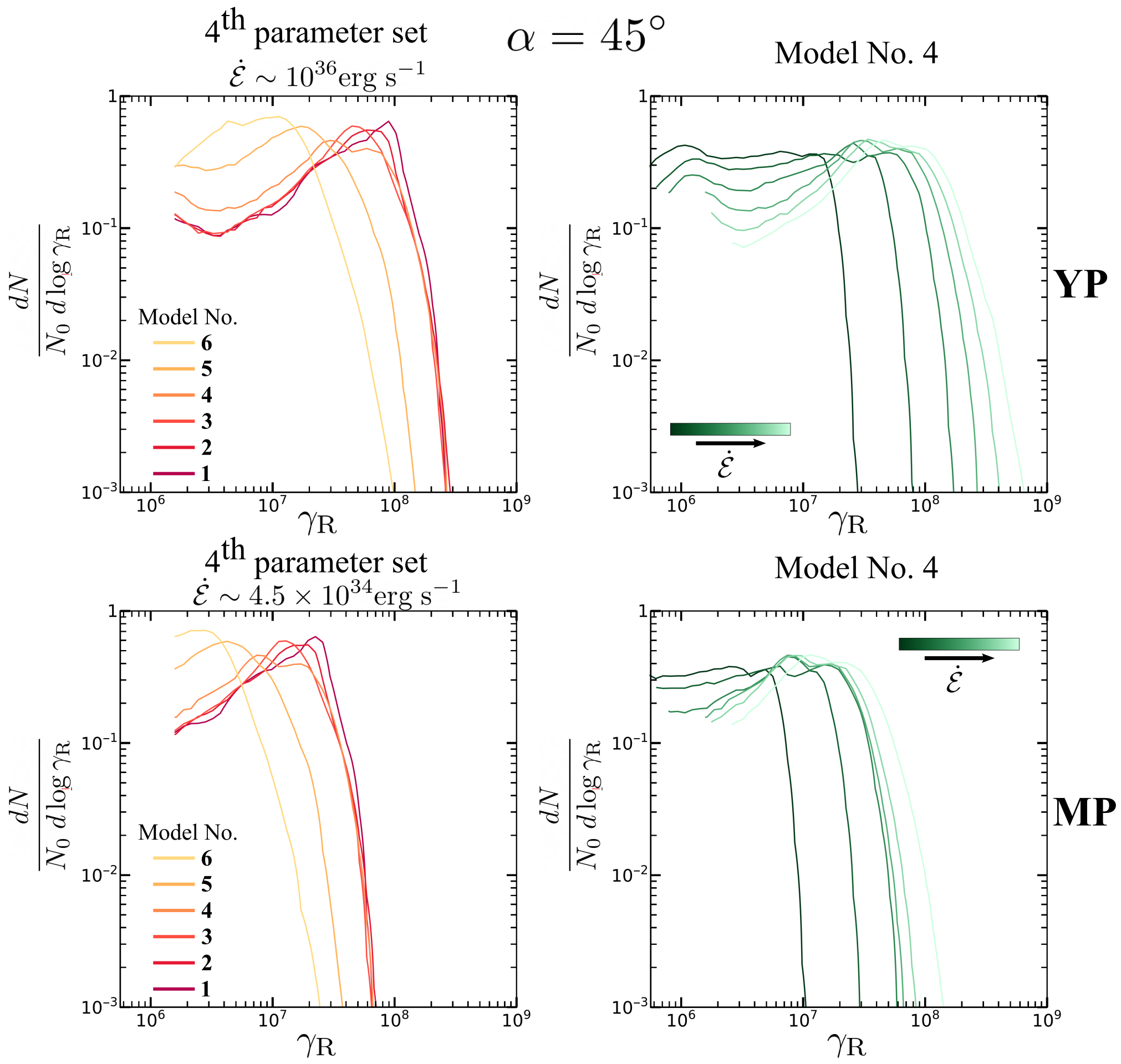}
  \end{center}
  \vspace{-0.2in}
  \caption{The physically realistic energy distributions of the high-energy particles for the \textit{separatrix zone} model and $\alpha=45^{\circ}$. The top (bottom) row shows YP (MP) models. The left-hand column shows distributions for the same $\ed$ value and different model numbers, i.e., different FF-ness. The right-hand column shows distributions for the same model number and for different $\ed$ values.}
  \label{fig:gammaR_distribution}
  \vspace{0.0in}
\end{figure*}

\begin{figure*}[!tbh]
\vspace{0.0in}
  \begin{center}
    \includegraphics[width=1.0\linewidth]{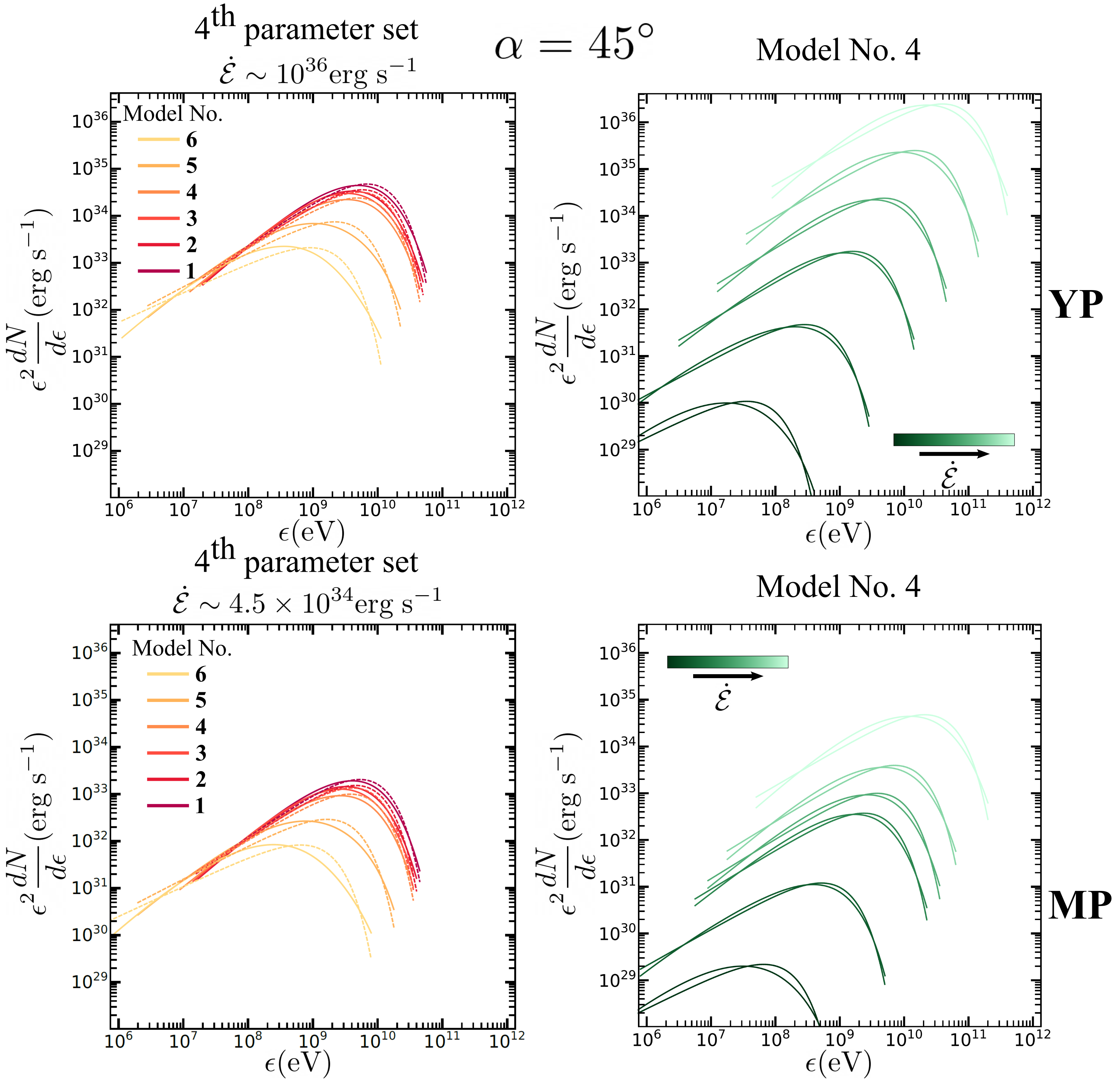}
  \end{center}
  \vspace{-0.2in}
  \caption{The model SED spectra are denoted by the solid lines for the cases shown in Figure~\ref{fig:gammaR_distribution}. The dashed lines denote the corresponding best fits assuming the model function in Eq.~\eqref{eq:spfit} for $b=1$.}
  \label{fig:spectra}
  \vspace{0.0in}
\end{figure*}

For each model, the energies of the
particles that encounter the high $\eacc$ are rescaled to realistic values.
For this, we follow the
procedure that is also described in \citetalias{2018ApJ...857...44K} assuming
the sets of realistic stellar surface magnetic field values
$B_{\star}$ and periods $P$ that are shown in
Table~\ref{tab:realpar}. These realistic value sets, which had been
also used in \citetalias{2018ApJ...857...44K} delineate the entire range
of spin-down powers of YPs and MPs.

\begin{table*}[!hbt]
\centering
        \begin{tabular}{cccccc}
            \hline
\multicolumn{3}{c|}{Young Pulsars} & \multicolumn{3}{c}{Millisecond Pulsars}\\
            \cline{1-6}\\[-8pt]
  $B_{\star}$ & $P$ & $\dot{\mathcal{E}}_{\rm FF}$ & $B_{\star}$ & $P$ &
$\dot{\mathcal{E}}_{\rm
FF}$ \\
              ($10^{12}$G) & (ms) & ($\rm erg\;s^{-1}$) & ($10^{8}$G) & (ms) & ($\rm
            erg\;s^{-1}$) \\
            \hline
 0.63 & 302.0 & $\sim 1\times 10^{33}$ & 0.7 & 5.1 & $\sim 1.5\times 10^{32}$\\
 1.26 & 239.9 & $\sim 1\times 10^{34}$ & 2.2 & 5.0 & $\sim 1.5\times 10^{33}$\\
 2.24 & 177.8 & $\sim 1\times 10^{35}$ & 5.0 & 4.1 & $\sim 2\times 10^{34}$\\
 3.16 & 398.1 & $\sim 1\times 10^{36}$ & 3.3 & 2.7 & $\sim 4.5\times 10^{34}$\\
 3.98 & 223.9 & $\sim 1\times 10^{37}$ & 3.5 & 2.0 & $\sim 1.5\times 10^{35}$\\
 7.94 & 125.9 & $\sim 1\times 10^{38}$ & 10.0 & 1.8 & $\sim 2\times 10^{36}$\\
\hline
        \end{tabular}
    \caption{The 12 (6+6) adopted realistic ($B_{\star},~P$) value sets for YP and MP
    models. For each value set, the corresponding FF spin-down power $\ed=4\pi^4r_{\star}^6B_{\star}^2(1+\sin^2\alpha)/(c^3P^4)$ is also indicated.}
    \label{tab:realpar}
\end{table*}

\section{Results} \label{sec:results}

\subsection{Energetics \& Spectra}\label{subsec:spectra}

Following \citetalias{2018ApJ...857...44K}, we identify as the actual accelerating
regions in the magnetosphere those that the ratio of the local accelerating
electric component, $\eacc$ over the local total electric field, $E$ exceeds
the value $10^{-1.1}$, i.e., $E_{\rm acc}\geq 10^{-1.1} E$\footnote{{These are the magnetosphere regions where the Ohmic dissipation $\mathbf{J}\cdot\mathbf{E}$ predominantly takes place. We note that $\eacc$ is the electric field, $E_0$, in the frame where $\mathbf{E}$ and $\mathbf{B}$ are parallel and is related to the Lorentz invariants through the relations $\mathbf{E}\cdot\mathbf{B}=E_0B_0$, $E^2-B^2=E_0^2-B_0^2$ \citep{2012arXiv1205.3367G,2012ApJ...746...60L}}}. In Figure~\ref{fig:002}, we plot the magnetic field
structures (blue streamlines) on the
$\pmb{\mu}-\pmb{\Omega}$\footnote{$\pmb{\mu}$ and $\pmb{\Omega}$ are the
stellar magnetic and angular momenta, respectively.} poloidal plane for the indicated
$\alpha$ values and models. The colored regions indicate the accelerating
regions according to the denoted color scale. These regions are mainly located around the reconnecting regions
beyond the LC and become progressively thinner towards the more FF models.

In Figure~\ref{fig:voltage_distribution}, we plot the distributions of the
potential drops, i.e., $V$ values, the particles encounter. The three panels
correspond to the indicated $\alpha$ values, while the different colors denote
the indicated model numbers. The highest $V$ values the
particles encounter decrease with an increasing model number, i.e., higher FF-ness. The black dashed line plotted in the
$\alpha=45^{\circ}$ panel corresponds to a simulation of $w=0.12$ and
$\ir_{\rm SZ}=1.5\ir_{\rm GJ}^0$. This line is close to the model No. 4 line,
which has been produced for a different set of $w$ and $\ir_{\rm SZ}$ values, i.e., $w=0.04$, and $\ir_{\rm SZ}=1.0\ir_{\rm GJ}^0$.
The model behavior depicted in Figure~\ref{fig:voltage_distribution} supports the qualitative description shown in the schematic of
Figure~\ref{fig:dissipative_character_schematic}.

In Figure~\ref{fig:gammaR_distribution}, we plot the distributions of the
rescaled realistic Lorentz factors, i.e., $\gamma_{\rm R}$, for
$\alpha=45^{\circ}$. The first and second rows show results for YP and MP
models, respectively. In the left-hand panels, the
plotted $\gamma_{\rm R}$ distributions correspond to the same $\ed$ value, i.e., $10^{36}\rm \, erg\;s^{-1}$ for YPs and $4.5\times 10^{34}\rm \, erg\;s^{-1}$ for MPs, but for different model numbers. The right-hand side panels show the $\gamma_{\rm R}$
distributions corresponding to model No. 4 but for different $\ed$ values denoted by the indicated color scheme.
The realistic particle energies depend on both the model FF-ness and the $\ed$ value. The $\gamma_{\rm R}$ values reach
$\gamma_{\rm R}\gtrsim 10^6$ for the lowest $\ed$ value and highest FF-ness and up to $\gamma_{\rm R}\gtrsim 10^8$ for the highest $\ed$ value and lowest FF-ness for YPs. The
corresponding $\gamma_{\rm R}$ range for MPs is smaller, which
reflects the shorter range of the corresponding $\ed$ values. For the same
model, i.e., same FF-ness, the $\gamma_{\rm R}$ values increase with $\ed$
because the absolute potential-drop values, i.e., absolute effective $\eacc$ values,
increase with $\ed$. Similarly, the energy of the high-energy particles increases for
the same $\ed$ value as the model FF-ness becomes weaker, i.e., higher potential-drop values. It is also evident that the particle energies are more
sensitive to $\ir_{\rm SZ}$ than to $w$. This implies that a considerable
increase of $w$ is required for a moderate increase of the corresponding potential drop.
In other words, the potential-drop values saturate, which is a good fraction of the entire potential drop in the polar cap, and struggle to increase more.

\begin{figure}[!tbh]
\vspace{0.0in}
  \begin{center}
    \includegraphics[width=1.0\linewidth]{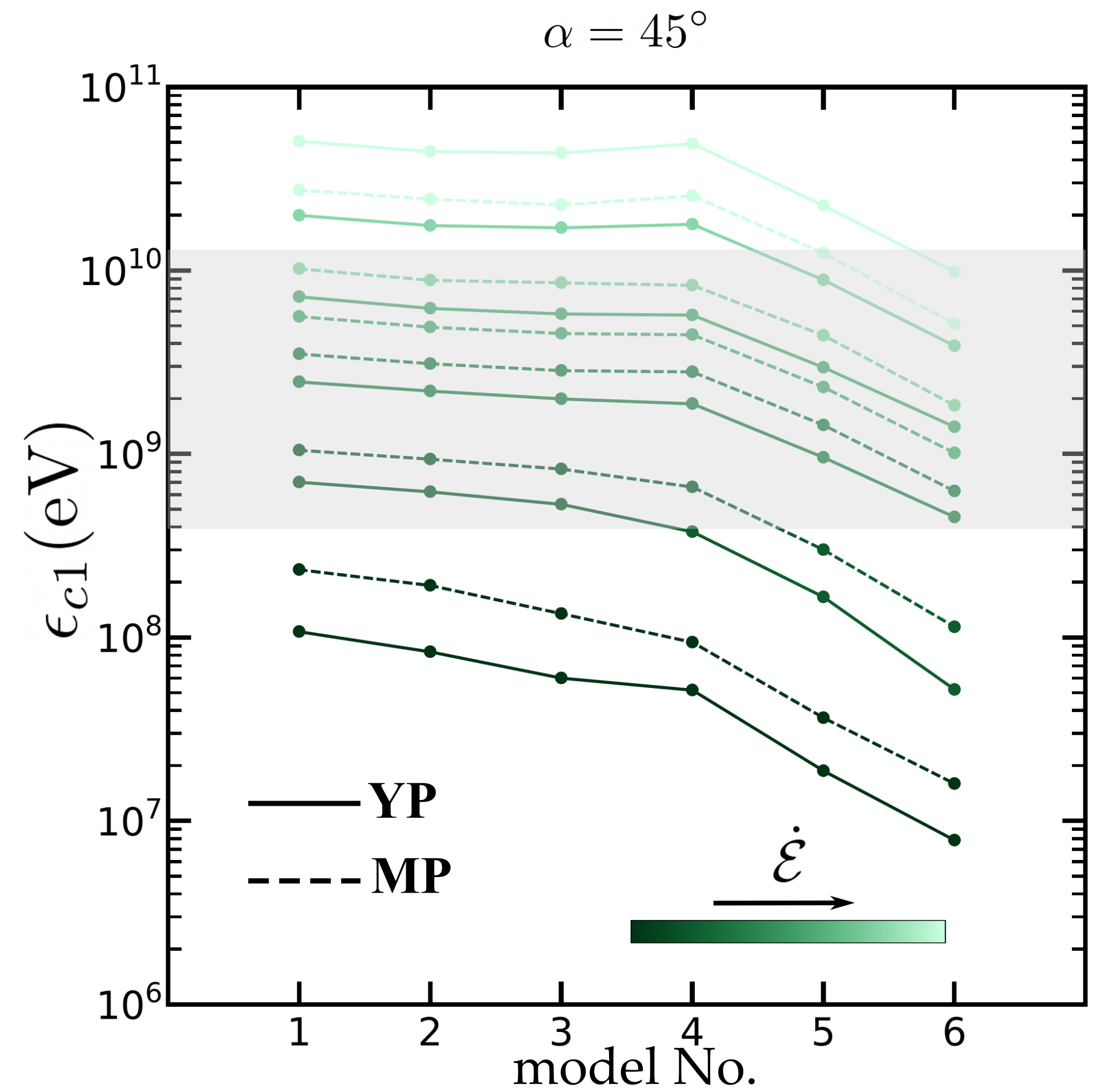}
  \end{center}
  \vspace{-0.2in}
  \caption{The $\ec$ values for the \textit{separatrix zone} models, $\alpha=45^{\circ}$ as a function of the model number. Each solid and dashed line corresponds to different YP and MP $\ed$ values, respectively. The indicated color scale denotes the increasing direction of $\ed$. Nonetheless, for clarity, the color scale has been adjusted differently for the YP and MP parameter sets in order to cover the corresponding entire $\ed$ range. The gray zone establishes the Fermi-compatible models.}
  \label{fig:ecut_model}
  \vspace{0.0in}
\end{figure}

The particle energies directly affect the corresponding $\gamma$-ray spectra. In
Figures~\ref{fig:spectra}, we plot the photon spectral energy distributions
(SED) corresponding to the rescaled particle energies shown in
Figure~\ref{fig:gammaR_distribution}. We note that
Figures~\ref{fig:gammaR_distribution} and \ref{fig:spectra} use the same
color notation, which allows a one-to-one comparison between the particle
energy distributions and the corresponding SED spectra. The solid lines show the superposition of the individual particle SEDs, which read, e.g., \citet{1998clel.book.....J}
\begin{equation}
    \label{eq:individual_spectrum}
    {\frac{d\mathcal{P}}{d\log\omega}=\frac{\sqrt{3}q_e^2\gamma_{\rm R}\omega_{\rm c}}{2\pi R_{\rm
    c}}\left(\frac{\omega}{\omega_{\rm
    c}}\right)^2\int_{\frac{\omega}{\omega_{\rm c}}}^{\infty}K_{5/3}(x)dx}
    \vspace{0.0in}
\end{equation}
where $\mathcal{P}$ the emitted power, $\omega$ the photon frequency, $R_{\rm c}$
the radius of curvature, $\omega_{\rm c}=3\gamma_{\rm R}c/(2R_{\rm c})$ the
critical frequency, and $K_{5/3}$ the modified Bessel function of the order
$5/3$. The dashed lines show the corresponding best fits assuming the model
function
\begin{equation}
    \label{eq:spfit}
    \frac{dN}{d\epsilon}=A \epsilon^{-\Gamma}\exp\left[-\left(\frac{\epsilon}{\epsilon_{\rm
    cut}}\right)^b\right]
    \vspace{0.0in}
\end{equation}
for $b=1$, where $A$ is the normalization factor, $\Gamma$ is the photon-index
and $\ec$ the cutoff energy. In \citet{2022ApJ...934...65K}, we showed that the cutoff energy corresponding to the pure exponential cutoff function model, i.e., Eq.~\eqref{eq:spfit} with $b=1$, optimally probes the maximum cutoff energy of the emission that originates from the core of the dissipative region, i.e., the ECS, which is mainly responsible for the emission around the peaks of the $\gamma$-ray light curves. Thus, even though the sub-exponential model function, i.e., $b\neq 1$, may better describe the totality of the spectra, especially for the more FF models, the cutoff energy that is more relevant to the $\ec$ in the FP relation is better captured by the one corresponding to the pure exponential model function, i.e., $b=1$ \citep[see][for more details]{2022ApJ...934...65K}. Below, adopting the nomenclature of \citet{2022ApJ...934...65K}, we denote $\epsilon_{\rm c1}$ the cutoff energy value corresponding to $b=1$.

\begin{figure*}[!tbh]
\vspace{0.0in}
  \begin{center}
    \includegraphics[width=1.0\linewidth]{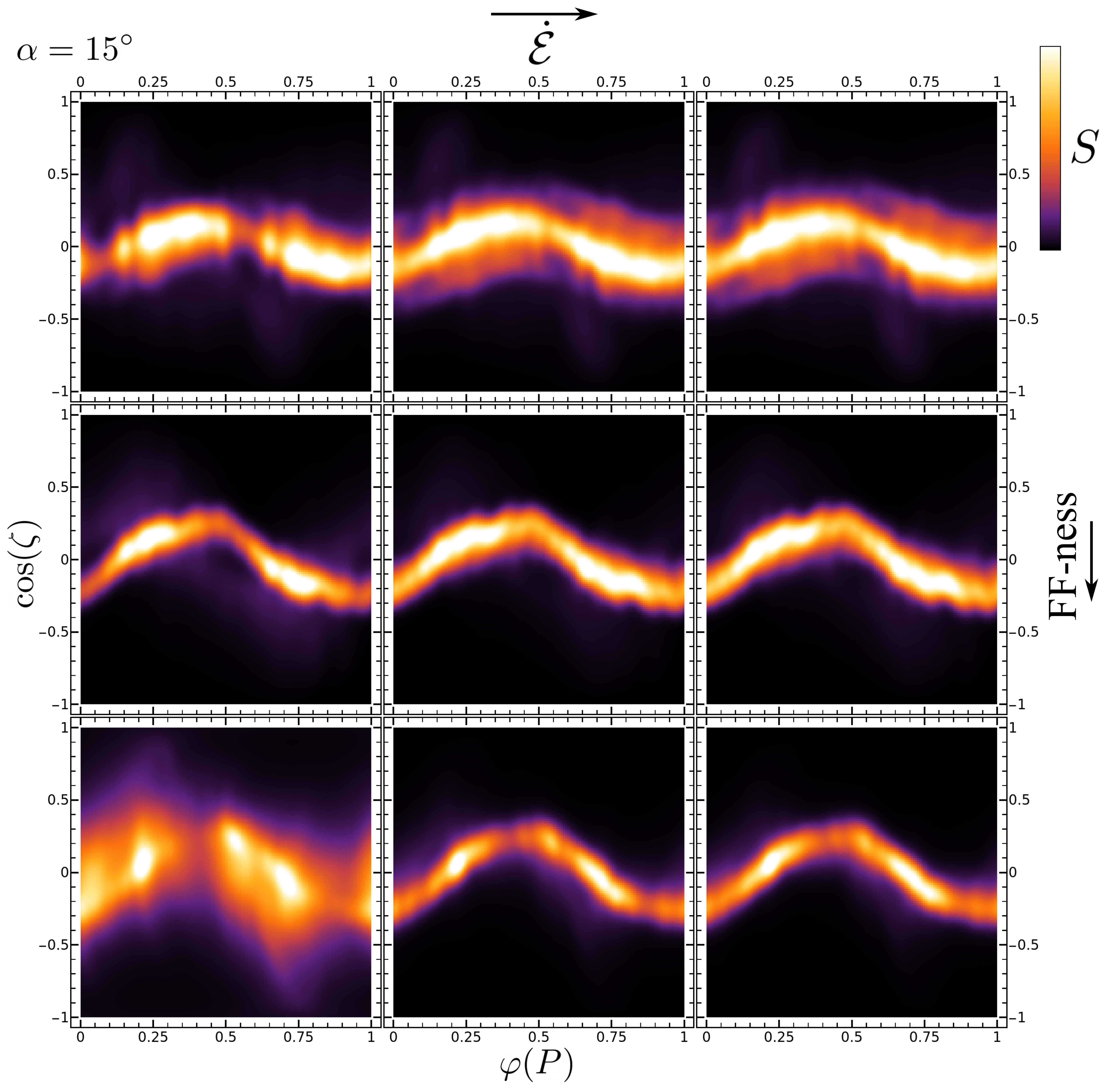}
  \end{center}
  \vspace{-0.2in}
  \caption{{The sky maps for the \textit{separatrix zone} model of YPs with $\alpha=15^{\circ}$. The $\ed$ increases from left to right while the FF-ness increases from top to bottom. More specifically, the plotted sky maps correspond to the model numbers 1, 4, and 6 and $\ed=10^{34},~10^{36}$, and $10^{38}\rm  erg\;s^{-1}$.} The color scale denotes the differential luminosity per steradian.}
  \label{fig:skymap15}
  \vspace{0.0in}
\end{figure*}

In Figure~\ref{fig:ecut_model}, we plot for YPs (solid lines) and for MPs
(dashed lines) the best fit $\epsilon_{\rm c1}$ parameter values as a function of the model number, for the $\alpha=45^{\circ}$ models. The different colors indicate different $\ed$ values as these are denoted in the figure. The $\ed$
values increase from the dark green lines towards the light green ones. The grayish horizontal zone denotes the $\epsilon_{\rm c1}$ range observed by Fermi-LAT. For high $\ed$, the $\epsilon_{\rm c1}$ values saturate very fast towards the less FF
models (model numbers smaller than 4). As shown below (Figure~\ref{fig:RRLR}), curvature radiation operates
closer to the RRL regime at high $\ed$ than at
low $\ed$ values. Therefore, the rather weak variation of the potential-drop values at low model numbers and the severe restrictions of the corresponding particle
energies (because of the RRL regime effect) results in very similar spectral $\epsilon_{\rm c1}$.

\begin{figure*}[!tbh]
\vspace{0.0in}
  \begin{center}
    \includegraphics[width=1.0\linewidth]{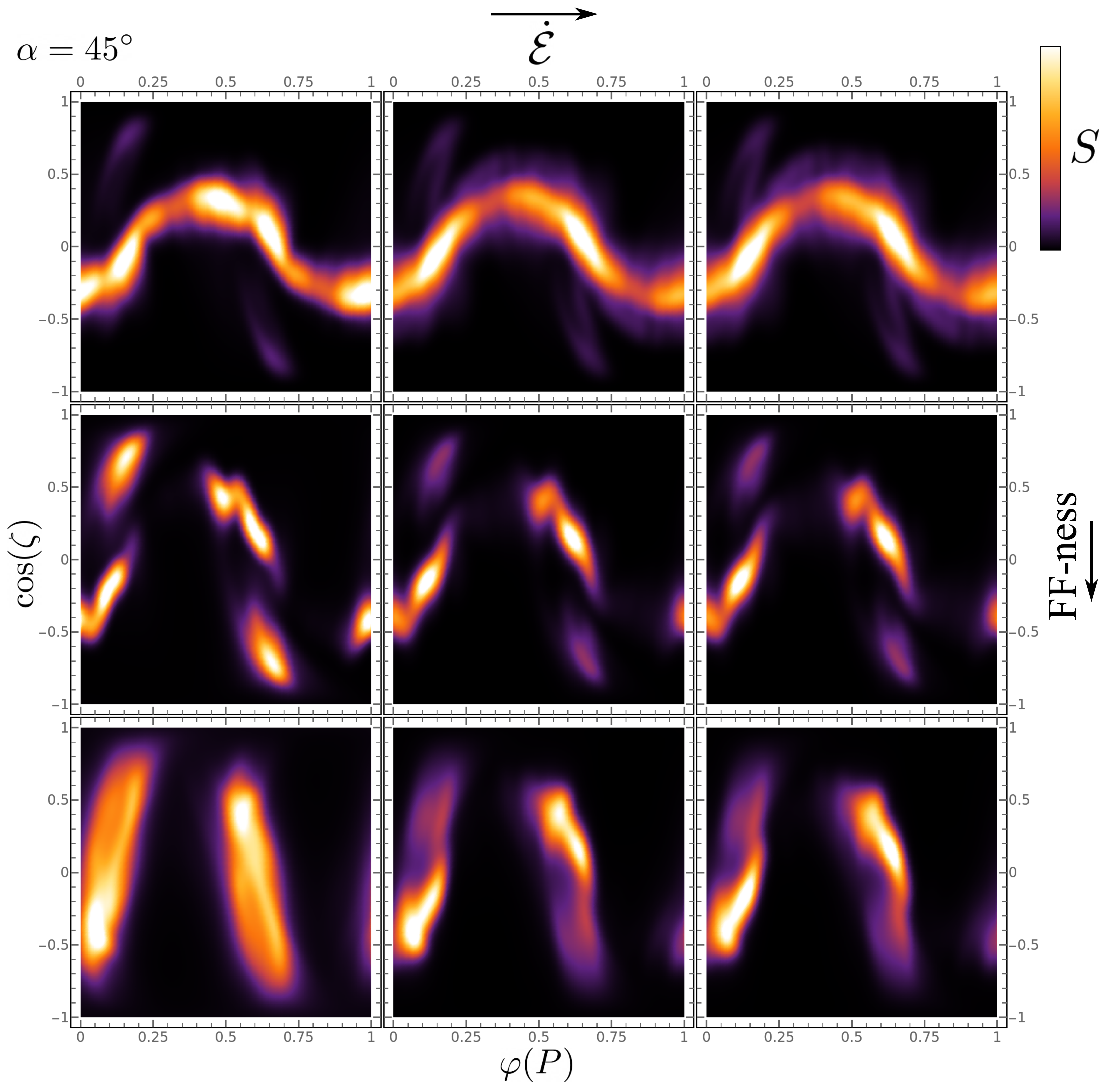}
  \end{center}
  \vspace{-0.2in}
  \caption{Similar to Figure~\ref{fig:skymap15} but for $\alpha=45^{\circ}$.}
  \label{fig:skymap45}
  \vspace{0.0in}
\end{figure*}

On the one hand, for the lowest considered $\ed$ values, which are slightly
below the $\ed$ values Fermi observes YPs and MPs, the $\epsilon_{\rm c1}$ values, even for
the least FF model, i.e., model No. 1, hardly exceed 100-200 MeV. On the other
hand, for the highest considered $\ed$ values, the most FF model, i.e., model No. 6, has $\epsilon_{\rm c1}$ that reaches below $\approx 10$ GeV. However, for the latter
case, it is apparent that higher particle injection rates in the separatrix zone would produce lower $\epsilon_{\rm c1}$ well within the range observed by Fermi. For
each $\ed$, i.e., realistic parameter set, we call Fermi-compatible or simply \emph{compatible} models those that
produce $\epsilon_{\rm c1}$ that lie within the Fermi corresponding values, i.e., within
the grayish zone in Figure~\ref{fig:ecut_model}. The model compatibility implies the FF-ness, i.e., the model number increases with $\ed$. Taking into account this remark, it becomes evident from Figure~\ref{fig:spectra} that the model spectra become wider with increasing FF-ness (maroon to yellow colored spectra), i.e., increasing $\ed$. A similar trend has been noticed in the Fermi spectra (D. Smith, private communication, 2022).

We note that the \textit{separatrix zone} models compared to the models
presented in \citetalias{2018ApJ...857...44K}, where the particle
injection rate was uniform along the different magnetic field lines,
have smaller maximum $\epsilon_{\rm c1}$ values for the lowest $\ed$ values because
the \textit{separatrix zone} models are less dissipative, i.e., higher FF-ness, and the corresponding
high-energy emission takes place always in the broader ECS region where the $E_{\rm acc}$ are smaller than those in regions inside the LC.
Moreover, for the \textit{separatrix zone} models, we managed, focusing
mainly on the particle injection rate in the separatrix zone, to produce
lower $\epsilon_{\rm c1}$ for the highest $\ed$ values than those for the same $\ed$ value in \citetalias{2018ApJ...857...44K}.

\begin{figure*}[!tbh]
\vspace{0.0in}
  \begin{center}
    \includegraphics[width=1.0\linewidth]{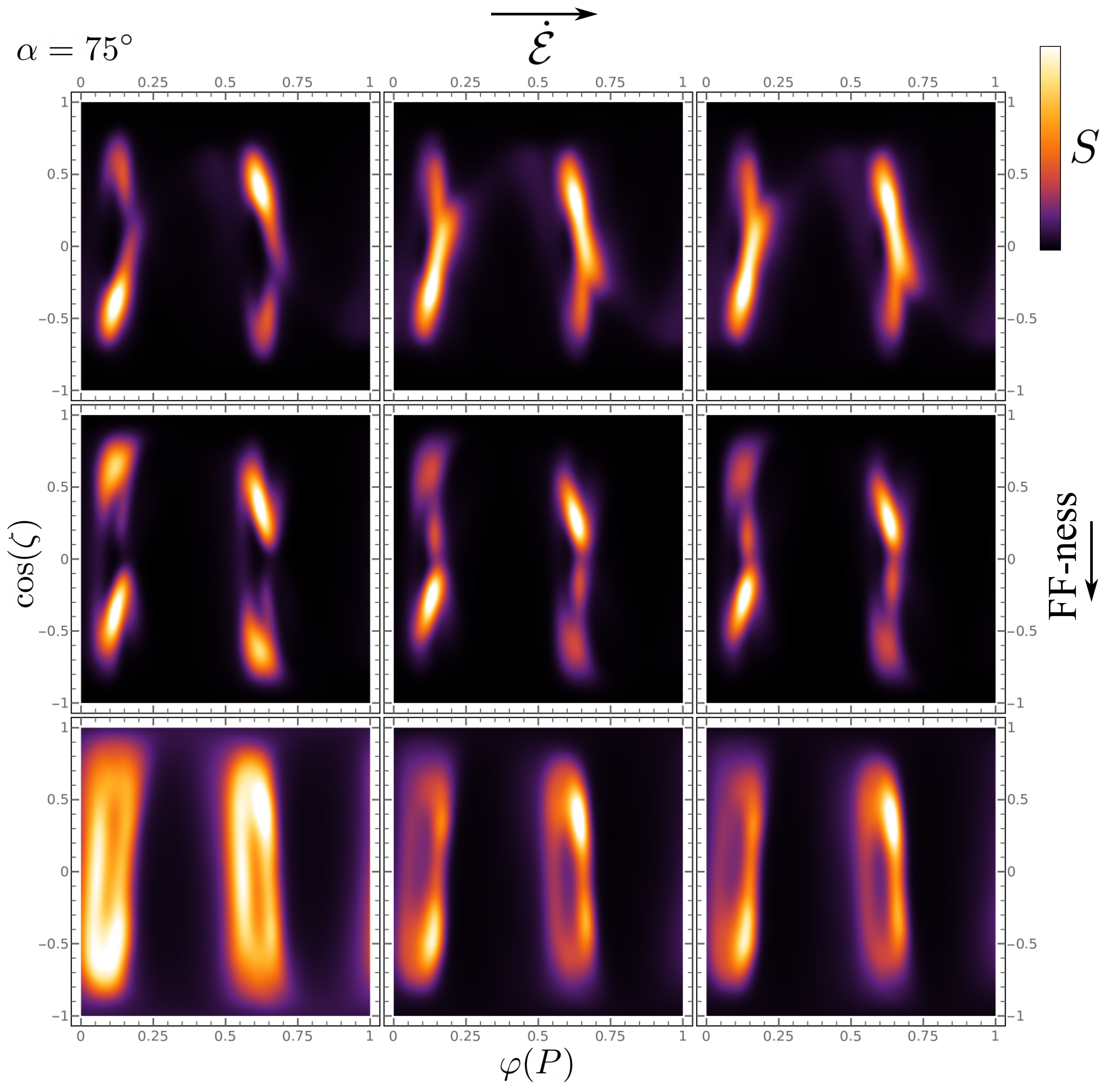}
  \end{center}
  \vspace{-0.2in}
  \caption{Similar to Figure~\ref{fig:skymap15} but for $\alpha=75^{\circ}$.}
  \label{fig:skymap75}
  \vspace{0.0in}
\end{figure*}

\subsection{Gamma-ray light curves}\label{subsec:gammaray_lightcurves}

Taking into account that the $\gamma$-ray photons are emitted along the
particle velocities, we calculate the bolometric luminosity per solid angle,
i.e., $S\equiv dL/d\Omega_{\rm s}$, distribution on the sky, the so-called sky map. In
Figures~\ref{fig:skymap15},~\ref{fig:skymap45}, and \ref{fig:skymap75}, we
present the model sky maps for the indicated $\ed$ and model number for
$\alpha=15^{\circ}$, $\alpha=45^{\circ}$, and $\alpha=75^{\circ}$,
respectively. The luminosity per solid angle is denoted in the indicated
color scale. In each panel, the horizontal axis depicts the rotational phase,
$\varphi$ while the vertical axis depicts the cosine of the observer angle, $\zeta$, which
is the angle between the rotational axis and the light of sight. We note that
$d\Omega_{\rm s}=\sin\zeta d\zeta d\varphi$. For the calculation of $\varphi$, the expression (18) of \citetalias{2018ApJ...857...44K} is used, which takes into
account the light travel time delays. Further details about the sky-map
calculation can be found in \citetalias{2018ApJ...857...44K}. Horizontal cross sections of
the sky maps at constant $\zeta$ provide the $\gamma$-ray
light curves observed along the corresponding line of sight.

The sky-map distributions in Figures~\ref{fig:skymap15}--\ref{fig:skymap75}
seem close to those corresponding to FIDO (FF inside dissipative outside) models \citep{2014ApJ...793...97K}
for all the models,
i.e., different model numbers, and for the entire range of $\ed$ values. This is an advancement compared to the sky maps corresponding
to the relatively low particle injection rate models of
\citetalias{2018ApJ...857...44K} (see the sky maps in the top row of Figures~15-17 of
\citetalias{2018ApJ...857...44K}). As mentioned above, the uniform particle injection rates along the
magnetic field lines in \citetalias{2018ApJ...857...44K}
resulted in sky maps, which for the relatively low $\ir$ values, were not
always consistent with the shape patterns of the Fermi-LAT $\gamma$-ray light
curves. We also note that for most FF models and for $\alpha=75^{\circ}$ ring type patterns start appearing in the sky maps, which are more prominent for the lowest $\ed$ values (see the bottom left panel in Figure~\ref{fig:skymap75}). We discuss this effect in detail in section \ref{sec:ecs}. In any case, we emphasize here that the sky maps of the compatible models (see
Figure~\ref{fig:ecut_model}) lie mainly along the top-left--bottom-right diagonal of
Figures~\ref{fig:skymap15}-\ref{fig:skymap75}, where this effect is weaker.

\begin{figure*}[!tbh]
\vspace{0.0in}
  \begin{center}
    \includegraphics[width=1.0\linewidth]{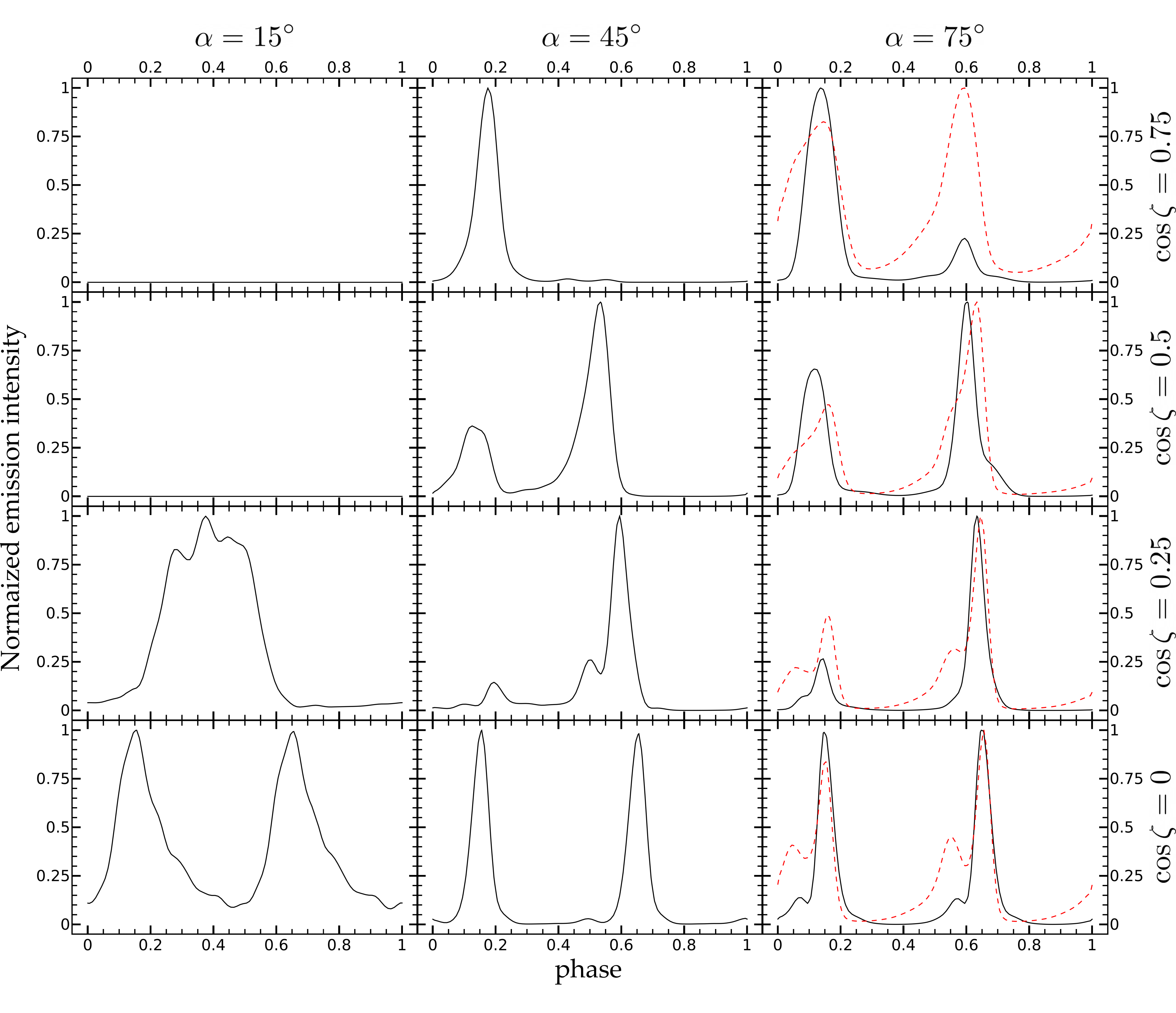}
  \end{center}
  \vspace{-0.2in}
  \caption{A $\gamma$-ray light curve atlas for the indicated $\alpha$ and $\cos\zeta$ values. The $\gamma$-ray light curves correspond to the middle panel sky-maps in Figures~\ref{fig:skymap15}-\ref{fig:skymap75}. The dashed red lines denote the $\gamma$-ray light curves corresponding to the bottom right panel of Figure~\ref{fig:skymap75}.}
  \label{fig:light curve atlas}
  \vspace{0.0in}
\end{figure*}

In Figure~\ref{fig:light curve atlas}, an atlas of YP $\gamma$-ray light curves corresponding to the middle sky-map panels of Figures~\ref{fig:skymap15}-\ref{fig:skymap75}, i.e, for $\ed=10^{36}\rm \, erg\;s^{-1}$, and model number 4, is plotted for demonstration. The different columns
correspond to the indicated $\alpha$ values, while the different rows
correspond to the indicated $\cos\zeta$ values. In the right-hand column, the red-dashed lines denote the $\gamma$-ray light curves corresponding to the ring-type sky maps plotted in the bottom right panel of Figure~\ref{fig:skymap75}. The shape patterns of the model
$\gamma$-ray light curves are similar to those observed by Fermi-LAT and
typically show no more than two peaks. Only the red light curves near $\cos\zeta=0$ exhibit weak and subdominant additional peaks.

The sky-map derivation allows the calculation of the corresponding beaming
correction factor, $f_{\rm b}(\zeta)$ (\citealt{2009ApJ...695.1289W,2010ApJ...714..810R}; \citetalias{2013ApJS..208...17A}) that
relates the observed energy flux at a certain distance and $\zeta$ value to the uniformly
distributed energy flux (over the different lines of sight, i.e., $\cos\zeta$) at the same distance. Thus, $f_{\rm b}$ reads
\begin{equation}
    \label{eq:beamingfactor1}
    f_{\rm b}(\zeta)=\frac{G_{\rm E}(d)}{G(d;\zeta)}
    \vspace{0.0in}
\end{equation}
where $G(d;\zeta)$ is the flux observed at some distance $d$ along $\zeta$
and $G_{\rm E}(d)=L_{\gamma}/4\pi d^2$ is the effective uniformly distributed
flux at $d$. Therefore, $f_{\rm b}$ along some $\zeta_0$ is calculated by
\begin{equation}
    \label{eq:beamingfactor2}
    f_{\rm b}(\zeta_0)=\frac{\int S(\varphi,\zeta)\sin \zeta d\zeta d\varphi}{2 \int
    S(\varphi,\zeta_0)
    d\varphi}\;.
    \vspace{0.0in}
\end{equation}
The beaming factor shows how representative the emission along a direction is compared to
total emission.

\begin{figure*}[!tbh]
\vspace{0.0in}
  \begin{center}
    \includegraphics[width=1.0\linewidth]{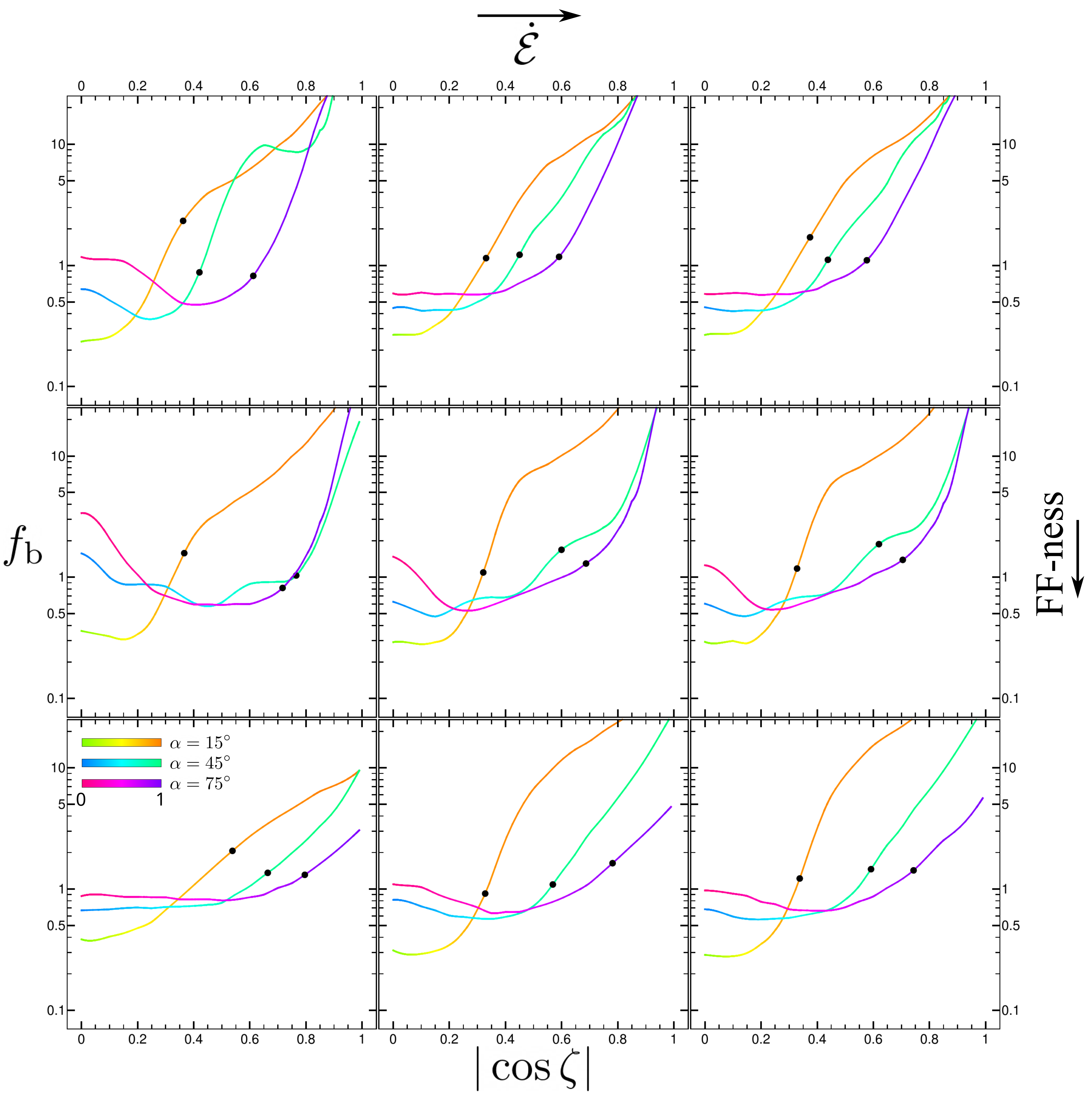}
  \end{center}
  \vspace{-0.2in}
  \caption{The beaming factors, $f_{\rm b}$ as a function of $\left| \cos\zeta \right|$ for all the sky maps that are plotted in Figures~\ref{fig:skymap15}-\ref{fig:skymap75}. There is a one-to-one correspondence between the panels plotted in this Figure and those of Figures~\ref{fig:skymap15}-\ref{fig:skymap75}. The color scales along the lines of different $\alpha$ values denote the fraction of the total $\gamma$-ray luminosity that is enclosed up to the corresponding $\left| \cos\zeta \right|$ value. The big black dots denote the $\left| \cos\zeta \right|$ value that encloses 95\% of the total $\gamma$-ray luminosity.}
  \label{fig:beamingfactor15}
  \vspace{0.0in}
\end{figure*}

In Figure~\ref{fig:beamingfactor15}, we plot $f_{\rm b}$ as a function of $\left|\cos\zeta \right|$\footnote{We note that the emission is symmetric with respect to the $\cos\zeta=0$ direction.}
for the cases corresponding to the sky maps that are plotted in Figures~\ref{fig:skymap15}-\ref{fig:skymap75}. In each panel, the different color scales correspond to the different $\alpha$ values, as these are denoted in the figure. The sky maps indicate that the emission is
mostly concentrated around the rotational equator, i.e., $\cos\zeta=0$.
Thus, the color along the lines denotes the fraction of the total emission
that is enclosed within the corresponding $\left|\cos\zeta \right|$ value. The black dots
on the lines indicate the $\cos\zeta$ values corresponding to 95\% of the
total emission. The $f_{\rm b}$ values are lower (higher) along the
directions, i.e., $\zeta$ values, where relatively more (less) emission is produced. We also see that $f_{\rm b}$ increases, in general, with $\alpha$
along the high emitting $\zeta$ values, i.e., $\cos\zeta$ close to 0.

Finally, we note that the $\varphi=0$ corresponds to the phase of a fiducial photon that originates from the magnetic pole at $r\rightarrow 0$. In
\citet{2014ApJ...793...97K}, we showed that assuming that the radio emission
originates from the magnetic pole at the stellar surface of YPs, the FIDO
model $\gamma$-ray light curves reproduce the observed $\delta-\Delta$
correlation \citepalias{2013ApJS..208...17A}. Using the compatible (see
Figure~\ref{fig:ecut_model}) model $\gamma$-ray light curves for the 3
different $\alpha$ values and for 20 different uniformly distributed
$\cos\zeta$ values, we identified the corresponding light-curve peaks, which
allowed the measurement of $\delta$ and $\Delta$ assuming that the
radio rotational phase corresponds to those waves that decouple proximate to the
magnetic poles at low altitudes.

The $\delta$ value is derived as the phase of the first peak of the $\gamma$-ray light curve (assuming 0 to be the phase emitted
at the magnetic pole on the stellar surface).
The $\Delta$ value is calculated as the phase difference (measured as a
fraction of the stellar period, $P$) between the two highest local
maxima. We note that any local maximum lower than 0.05 of
the value of the highest (first) maximum is not counted as a
local maximum and the corresponding light curve is considered to have only one peak. For the computation of the $\delta,~\Delta$ values, we use the algorithm we developed in \citet{2014ApJ...793...97K}. This algorithm takes into account most of the human-eye criteria employed in the determination of ($\delta$, $\Delta$) values, which allows an unbiased and fast computation of the ($\delta$, $\Delta$) values. Such an automated
calculation may lead to some erroneous values. However, the
expected number of such erroneous points is small since we have checked tens of light curves of various types, and in
the vast majority of the cases, the results were in total agreement
with those derived by simple visual inspection.

\begin{figure}[!tbh]
\vspace{0.0in}
  \begin{center}
    \includegraphics[width=1.0\linewidth]{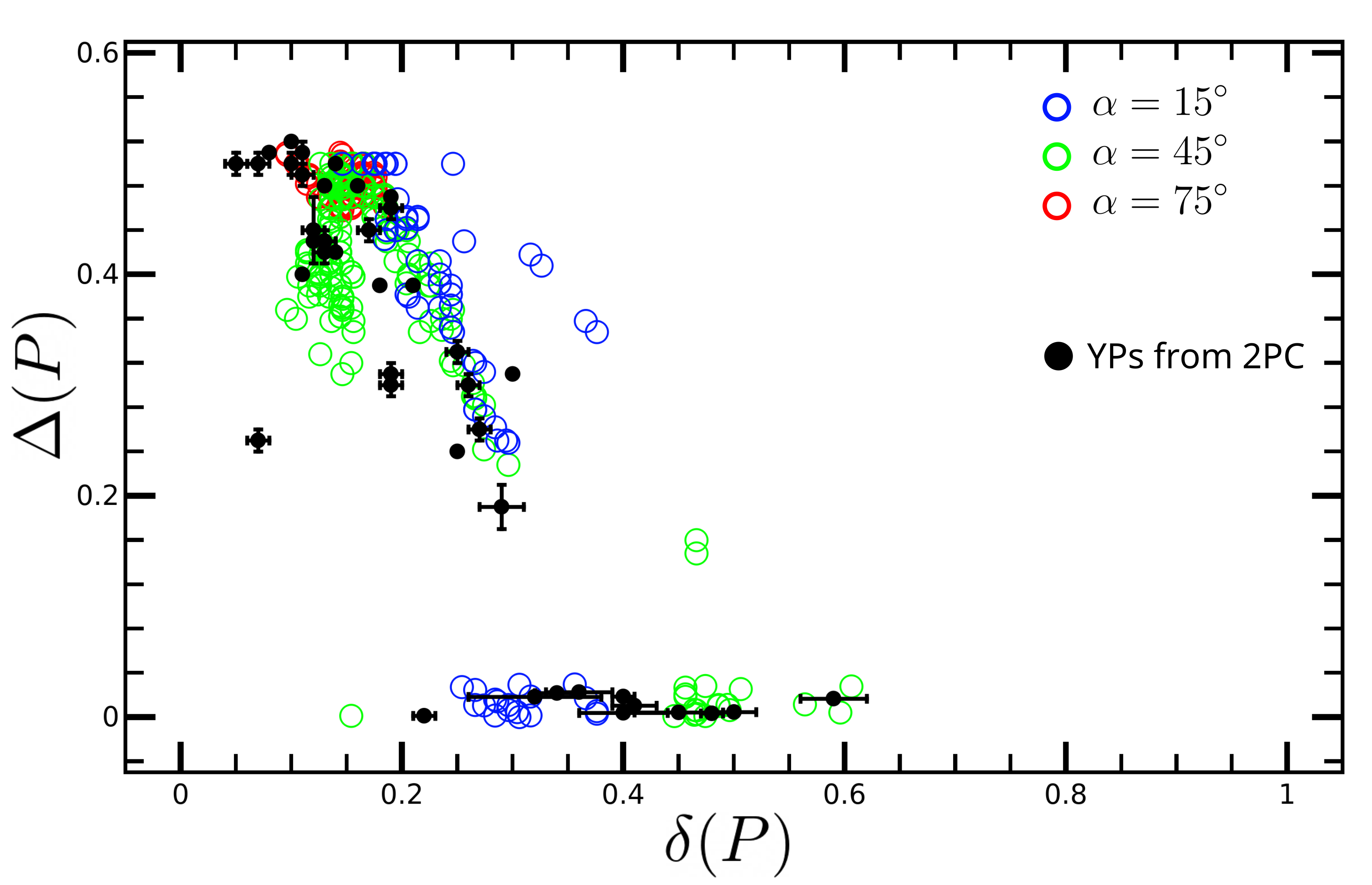}
  \end{center}
  \vspace{-0.2in}
  \caption{The model $\Delta-\delta$ values of YPs (denoted by the colored open circles) together with the corresponding observed ones from the \citetalias{2013ApJS..208...17A} (denoted by the black points with error bars).
  Following \citetalias{2013ApJS..208...17A}'s convention, we artificially stagger the model points along the horizontal axis, i.e., $\Delta=0$ to enhance clarity.}
  \label{fig:delta_Delta}
  \vspace{0.0in}
\end{figure}

In Figure~\ref{fig:delta_Delta}, we plot the model $(\delta,~\Delta)$ pairs denoted by
colored points together with the corresponding
\citetalias{2013ApJS..208...17A} black points. We see that the model
points reproduce the correlation established in
\citetalias{2013ApJS..208...17A} for YPs remarkably well. A more detailed comparison would require the incorporation of a higher number of
$\alpha$ values. In Figure~\ref{fig:delta_Delta}, we have considered that
radio pulsations are always observed\footnote{This seems to be true for high
$\ed$ values (see Figure 20 in \citetalias{2013ApJS..208...17A}).}. However,
this may not be the case for all the model points presented in
Figure~\ref{fig:delta_Delta}, especially for the $\alpha=15^{\circ}$ models
in which the $\zeta$ values along which most of the $\gamma$-rays are emitted
considerably differ from the $\zeta=15^{\circ}$ values corresponding to the
magnetic pole. Nonetheless, the model points in Figure~\ref{fig:delta_Delta} still
demonstrate that the $\gamma$ pulsation morphology (captured by $\Delta$) and the phase
differences between the radio, i.e., emission near magnetic poles, and the
$\gamma$-ray emission, i.e., $\delta$, is correlated in our models.

\subsection{Fundamental Plane}\label{sec:fundamental plane}

\begin{figure*}[!tbh]
\vspace{0.0in}
  \begin{center}
    \includegraphics[width=1.0\linewidth]{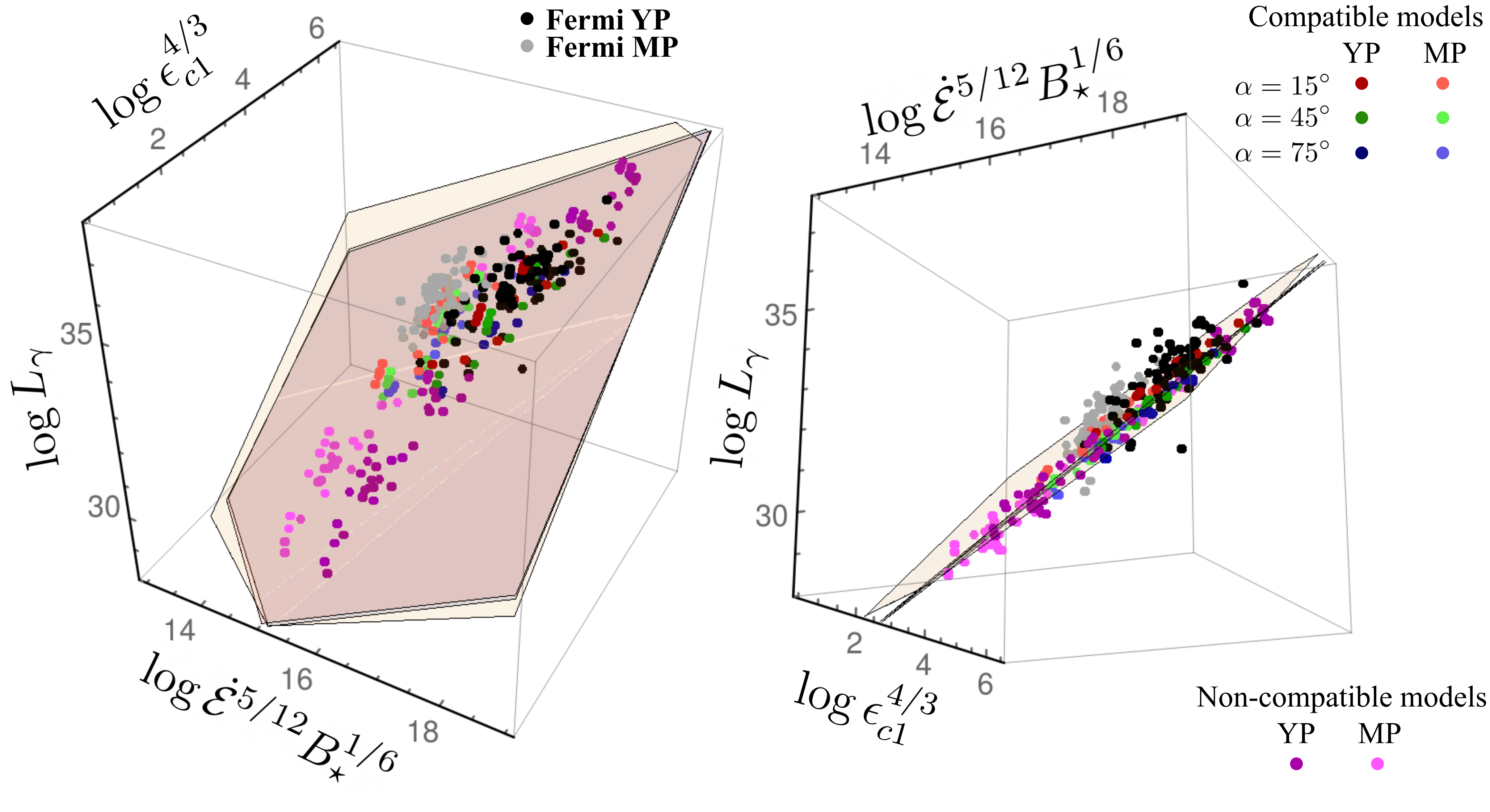}
  \end{center}
  \vspace{-0.2in}
  \caption{The FP of the model and Fermi-LAT objects are seen from different points of view in the reduced dimensionality space. The point color code is indicated in the Figure. The incompatible models, i.e., magenta-colored points, lie on the FP, but their $\ec$ are not within the observed by Fermi-LAT range (see Figure~\ref{fig:ecut_model}). The bluish and reddish planes denote the best model and 4FGL fits, i.e., Eqs.~\eqref{eq:fp_all_pic} and \eqref{eq:fp_4FGL}, respectively. We also note that the $\epsilon_{\rm c1}$ values for both the Fermi pulsars and the models correspond to the spectral fittings with $b=1$ \citep[see Eq.~(\ref{eq:spfit}) and][]{2022ApJ...934...65K}.}
  \label{fig:FP4FGLandPICALL}
  \vspace{0.0in}
\end{figure*}

\begin{figure}[!tbh]
\vspace{0.1in}
  \begin{center}
    \includegraphics[width=1.0\linewidth]{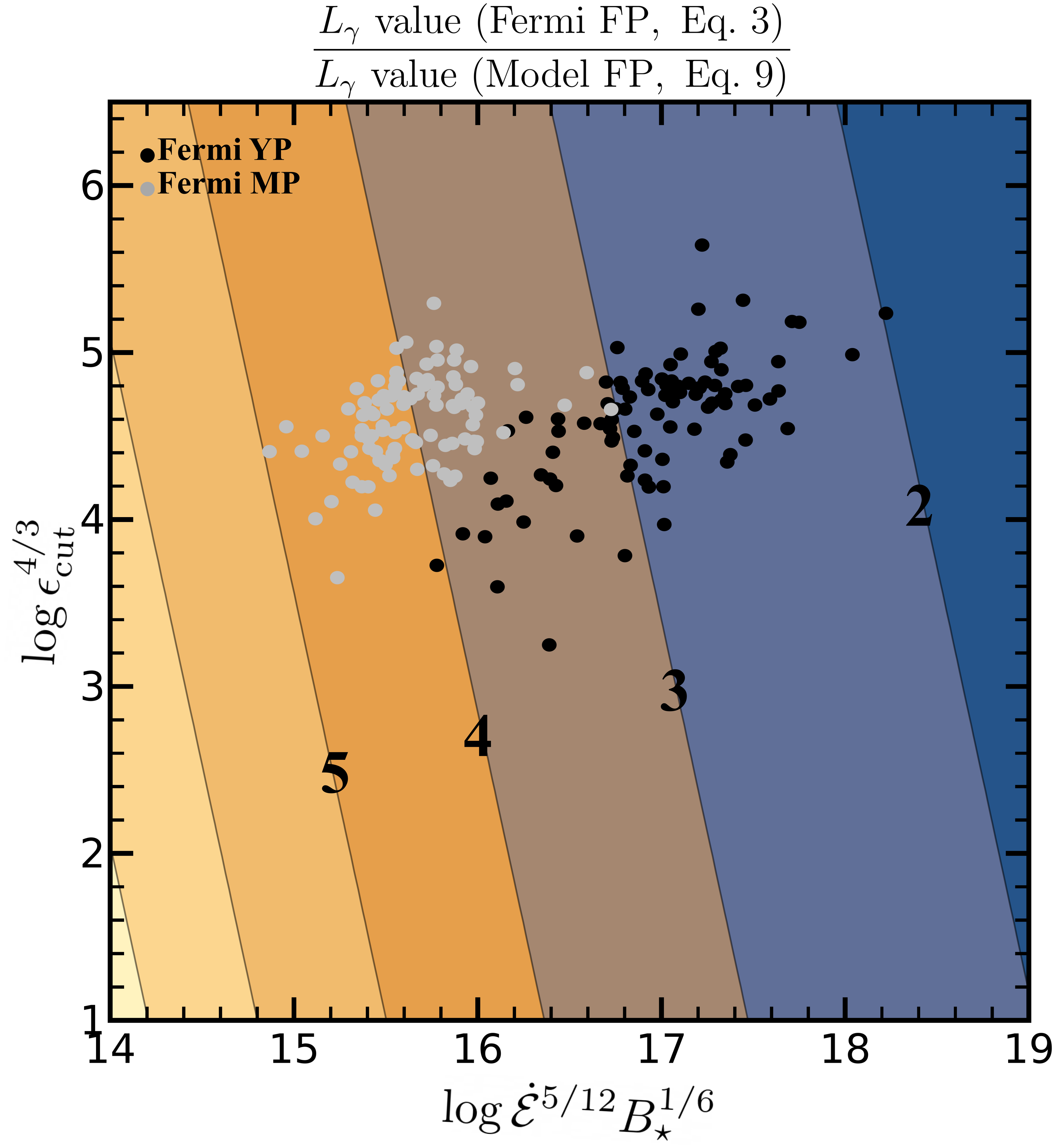}
  \end{center}
  \vspace{-0.2in}
  \caption{The contour plot of the $L_{\gamma}$ ratio corresponding to the 4FGL $L_{\gamma}$, i.e., Eq.~\eqref{eq:fp_4FGL} over the model compatible $L_{\gamma}$, i.e., Eq.~(\ref{eq:fp_compatible_pic}). The plotted 4FGL YPs and MPs mark the region where the actual objects lie.}
  \label{fig:FPprojected4FGLandRatioContours}
  \vspace{0.0in}
\end{figure}

Our PIC models allow the calculation of not only the emitted spectra but also the calculation of the bolometric $L_{\gamma}$ values, which (for each model) are simply the sum over all the PIC macroparticles of $2q_e^2\gamma_{\rm R}^4 c/3R_{\rm C}^2$.
The total number of PIC models is $216(=3\;\alpha\;{\rm values}\times
6 \;(\ir_{\rm SZ},~w)\; {\rm value\; sets} \times 12\; \ed\; {\rm values})$
while the number of the compatible PIC models (see
Figure~\ref{fig:ecut_model}) is 110.

With the model $\ed,~B_{\star},~\epsilon_{\rm c1}$, and $L_{\gamma}$ values at hand, we can calculate the model FP relations. Thus, the FP corresponding to the total number of PIC models reads
\begin{equation}
\label{eq:fp_all_pic} L_{\gamma}= 10^{12.6\pm 0.7}\epsilon_{\rm c1}^{1.39\pm
0.06}B_{\star}^{0.17\pm 0.02}\ed^{0.41\pm 0.04}
\end{equation}
while the FP corresponding only to the \emph{compatible} PIC models reads
\begin{equation}
\label{eq:fp_compatible_pic} L_{\gamma}= 10^{14.0\pm 1.1}\epsilon_{\rm c1}^{1.59\pm
0.13}B_{\star}^{0.18\pm 0.02}\ed^{0.34\pm 0.05}
\end{equation}
which are close to each other, both reproducing the
theoretically predicted dependencies. Following \citet{2022ApJ...934...65K}, we reduce the FP dimensionality, i.e., from 4 to 3, by defining 3 new variables $(\mathcal{X},~\mathcal{Y},~\mathcal{Z})=(\ed^{5/12}B_{\star}^{1/6},~\epsilon_{\rm c1}^{4/3},~L_{\gamma})$.
In Figure~\ref{fig:FP4FGLandPICALL}, we present, in the $\mathcal{(X,~Y,~Z)}$ space, the observed sample of 190 Fermi pulsars \citep{2022ApJ...934...65K}
together with the PIC models. The black and gray points denote the observed YPs
and MPs, respectively. The dark (light) red, green, and blue points denote the
$\alpha=0^{\circ}$, $\alpha=45^{\circ}$, and $\alpha=75^{\circ}$ PIC
compatible YP (MP) models. The dark and light magenta points denote the YPs and MPs,
respectively, of the rest, i.e., non-compatible PIC models. The two plotted FPs
are the observed one (transparent
red color, see Eq.~\ref{eq:fp_4FGL}) and the PIC one (transparent blue, see Eq.~\ref{eq:fp_all_pic}).

The FP relations
\eqref{eq:fp_4FGL}, \eqref{eq:fp_all_pic}, and
\eqref{eq:fp_compatible_pic} indicate that the model $L_{\gamma}$ values are
lower than the reported observed ones, even though the dependencies between the various
parameters are similar and always consistent with the predicted theoretical FP of
Eq.~(\ref{eq:fp_theory}). In
Figure~\ref{fig:FPprojected4FGLandRatioContours}, we compare the $L_{\gamma}$
values corresponding to the observed pulsars, i.e.,
Eq.~(\ref{eq:fp_4FGL}), and the compatible PIC models, i.e.,
Eq.~(\ref{eq:fp_compatible_pic}). The contour lines and shaded areas, which are plotted on the $(\mathcal{X},~\mathcal{Y})$ plane, indicate the different
$L_{\gamma_{\rm obs}}/L_{\gamma_{\rm opt}}$-ratio values, where
$L_{\gamma_{\rm obs}}$ and $L_{\gamma_{\rm opt}}$ are given by
Eqs.~\eqref{eq:fp_4FGL} and \eqref{eq:fp_compatible_pic}, respectively. The
large points are the observed pulsars, which indicate the observed area of
the FP. We see that in the \emph{observed}
area of the FP, $L_{\gamma_{\rm obs}}$ is 2 to 5 times higher than
$L_{\gamma_{\rm opt}}$. Nonetheless, at least a significant part of this difference is consistent with the model $f_{\rm b}$ values shown in
Figure~\ref{fig:beamingfactor15}, which indicates
that the average $f_{\rm b}$ values, especially for the high intensity
$\zeta$ values are less than 1, reaching for low $\alpha$ values lower than 0.5. According to the PIC
models, the $L_{\gamma_{\rm obs}}$ values, which have been calculated assuming
$f_{\rm b}=1$ are actually overestimations of the corresponding true values implying that the difference between the observations and models depicted in Figure~\ref{fig:FPprojected4FGLandRatioContours} is actually smaller.

\begin{figure*}[!tbh]
\vspace{0.0in}
  \begin{center}
    \includegraphics[width=1.0\linewidth]{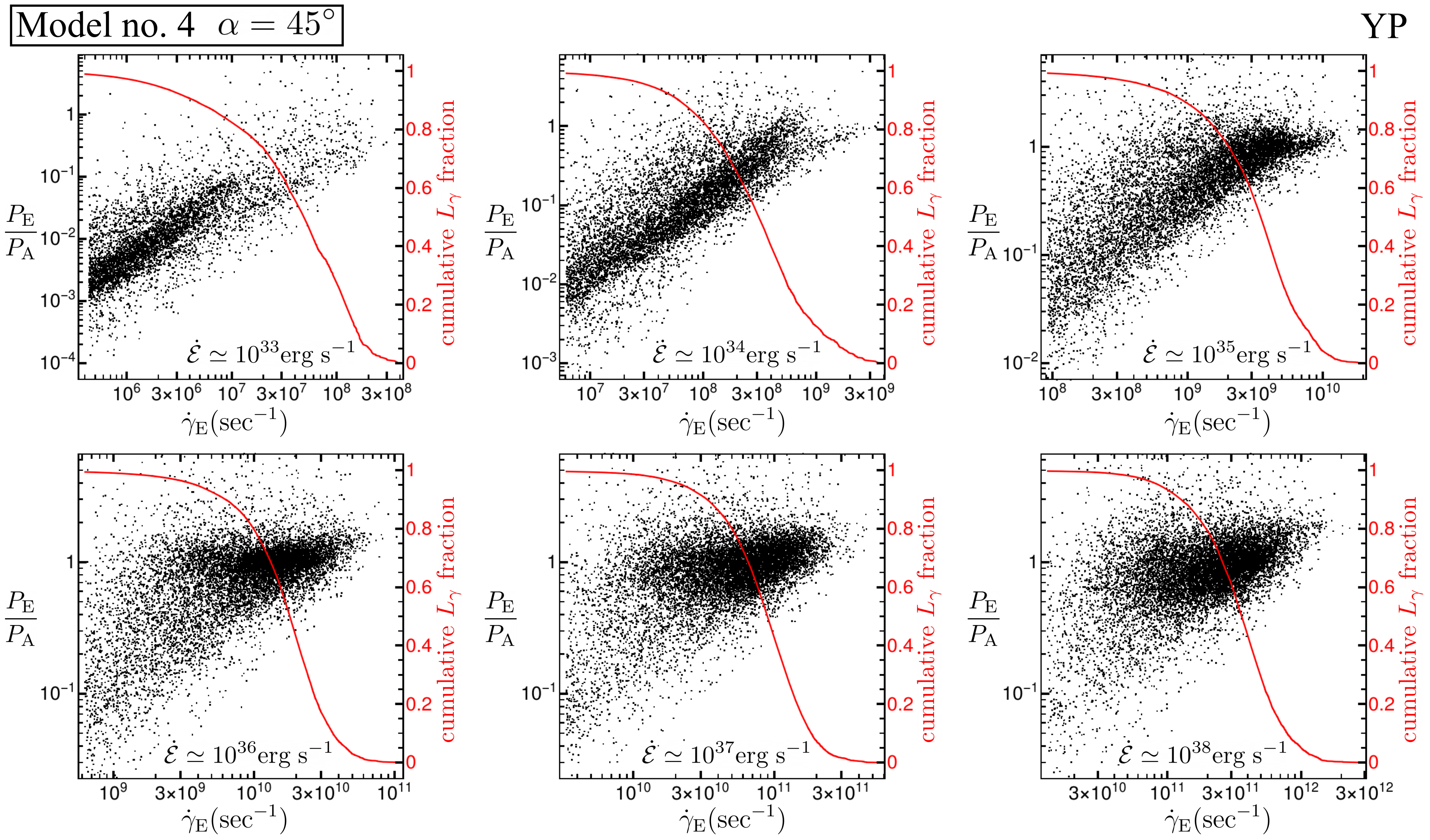}
  \end{center}
  \vspace{-0.2in}
  \caption{The ratio of the emitting power over the accelerating power as a function of the rate of radiation reaction losses. All the panels correspond to the \textit{separatrix zone} model number 4 for $\alpha=45^{\circ}$. Each panel shows the simulation particles for the indicated $\ed$ values of YPs. The red lines indicate the fraction of the total $\gamma$-ray luminosity that is emitted by all the particles with $\geq\dot{\gamma}_{\rm E}$ as this is depicted in the right-hand vertical axes. The high emitting particles do not reach the RRL regime, i.e., $P_{\rm E}/P_{\rm A}\approx 1$ for $\ed\lesssim 10^{34}\rm erg\;s^{-1}$.}
  \label{fig:RRLR}
  \vspace{0.0in}
\end{figure*}

\begin{figure*}[!tbh]
\vspace{0.0in}
  \begin{center}
    \includegraphics[width=1.0\linewidth]{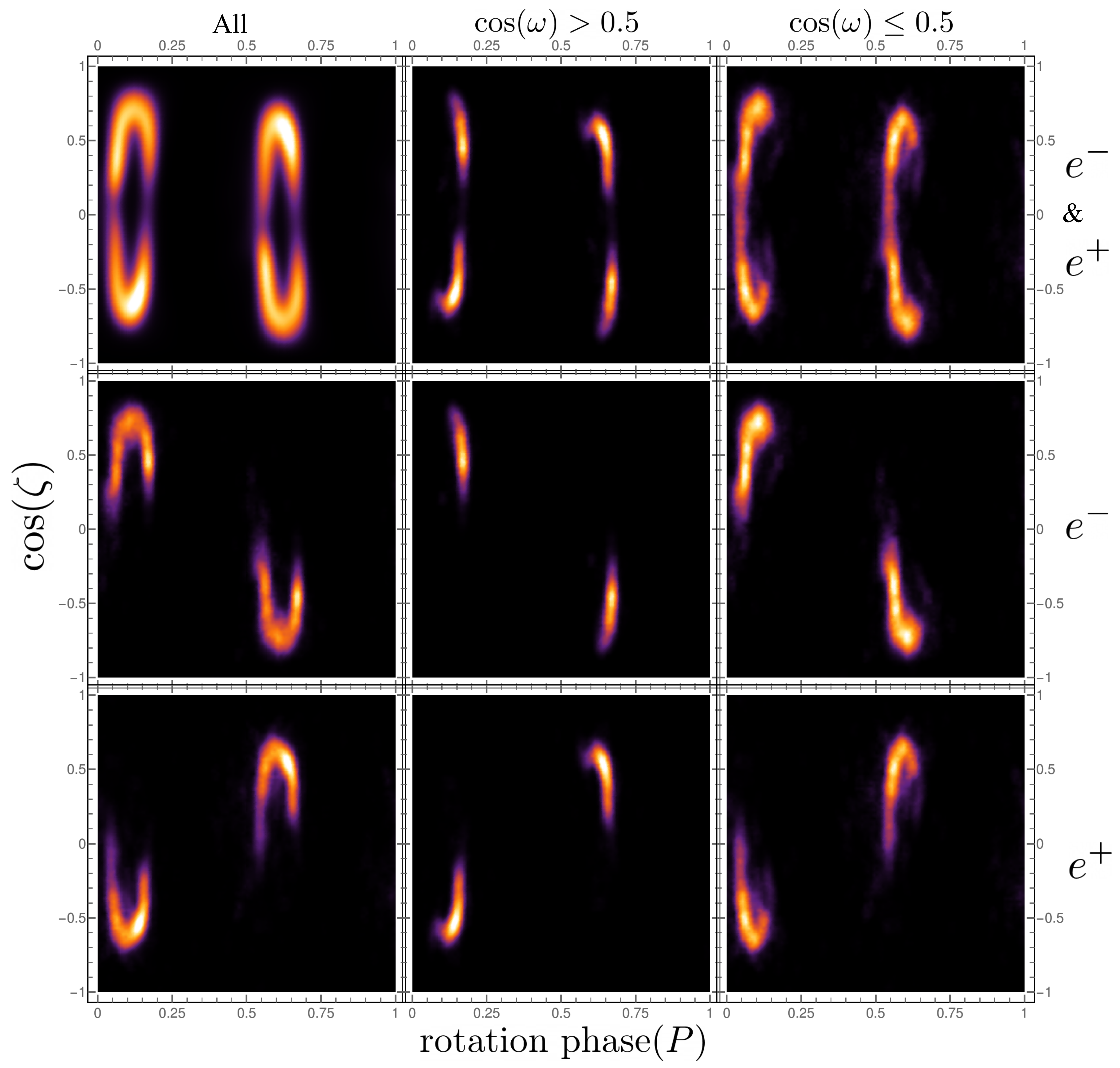}
  \end{center}
  \vspace{-0.2in}
  \caption{The total sky map (top, left-hand panel) for the \textit{extended separatrix zone} model for $\alpha=75^{\circ}$ and $\ed\approx 10^{38}\rm erg\; s^{-1}$ together with various particle sky-map components. The particle in the middle column, i.e., $\cos\omega>0.5$, are mainly located outside the LC, while the particle in the right-hand column, i.e., $\cos\omega\leq0.5$, are mainly located outside the LC. The middle and bottom rows show sky-map components corresponding to $e^-$ and $e^+$, respectively.}
  \label{fig:eSI_sky-maps_ele-pos_in-out}
  \vspace{0.0in}
\end{figure*}

\subsubsection{The applicability of the RRL regime}\label{sec:RRL_efficiancy}

The original theoretical derivation of the FP of $\gamma$-ray pulsars was motivated by the assumption that the $\gamma$-ray emission always operates in the RRL regime \citep{2019ApJ...883L...4K}. However, it was realized that the RRL emission is a sufficient but not necessary condition for the derivation of the FP. Figure~\ref{fig:FP4FGLandPICALL} and Eqs.~\eqref{eq:fp_all_pic},~\eqref{eq:fp_compatible_pic} demonstrate that our PIC models lie on the FP. Below, we show the applicability of the RRL regime in the PIC models and, more specifically, its variation with $\ed$.

In Figure~\ref{fig:RRLR}, we plot, for $\alpha=45^{\circ}$ and model number 4, the ratio of the particle energy-loss rates, i.e., emission power, $P_{\rm E}$, over the particle energy-gain rates, i.e., acceleration power, $P_{\rm A}$, as a function of the corresponding energy-loss rates, expressed in the corresponding $\dot{\gamma}_{\rm E}$. Each panel corresponds to the indicated six $\ed$ values for YPs. The red solid lines denote the cumulative fraction, shown in the right-hand vertical axis, of the emitting power from all the particles with emission power greater than the corresponding $\dot{\gamma}_{\rm E}$, i.e., $P_{\rm E}=\dot{\gamma}_{\rm E}m c^2$, value. We see that for $\ed\gtrsim 10^{34}\rm \, erg\;s^{-1}$ the vast majority of the particles and especially those with the highest emission powers emit in the RRL regime, i.e., $P_{\rm E}/P_{\rm A}\approx 1$. For lower $\ed$ values, it becomes  evident that even the high-power particles struggle to reach the RRL regime. This, which is also true for model number 1, i.e., the least FF models, is essentially the result of the corresponding lower absolute $\eacc$ values, i.e., potential drops, at the broader ECS region. We also note that for MPs the highest-energy particles start struggling to reach the RRL regime for $\ed\lesssim 10^{33}\rm erg\;s^{-1}$. Therefore, our results indicate that the non-RRL emission starts for $\ed$ values slightly below the currently observed ones, which is proximate to the pulsar $\gamma$-ray death line. Implications of such a transition and its detailed relevance to the pulsar $\gamma$-ray death line are deferred to a forthcoming publication.

\begin{figure*}[!tbh]
\vspace{0.0in}
  \begin{center}
    \includegraphics[width=1.0\linewidth]{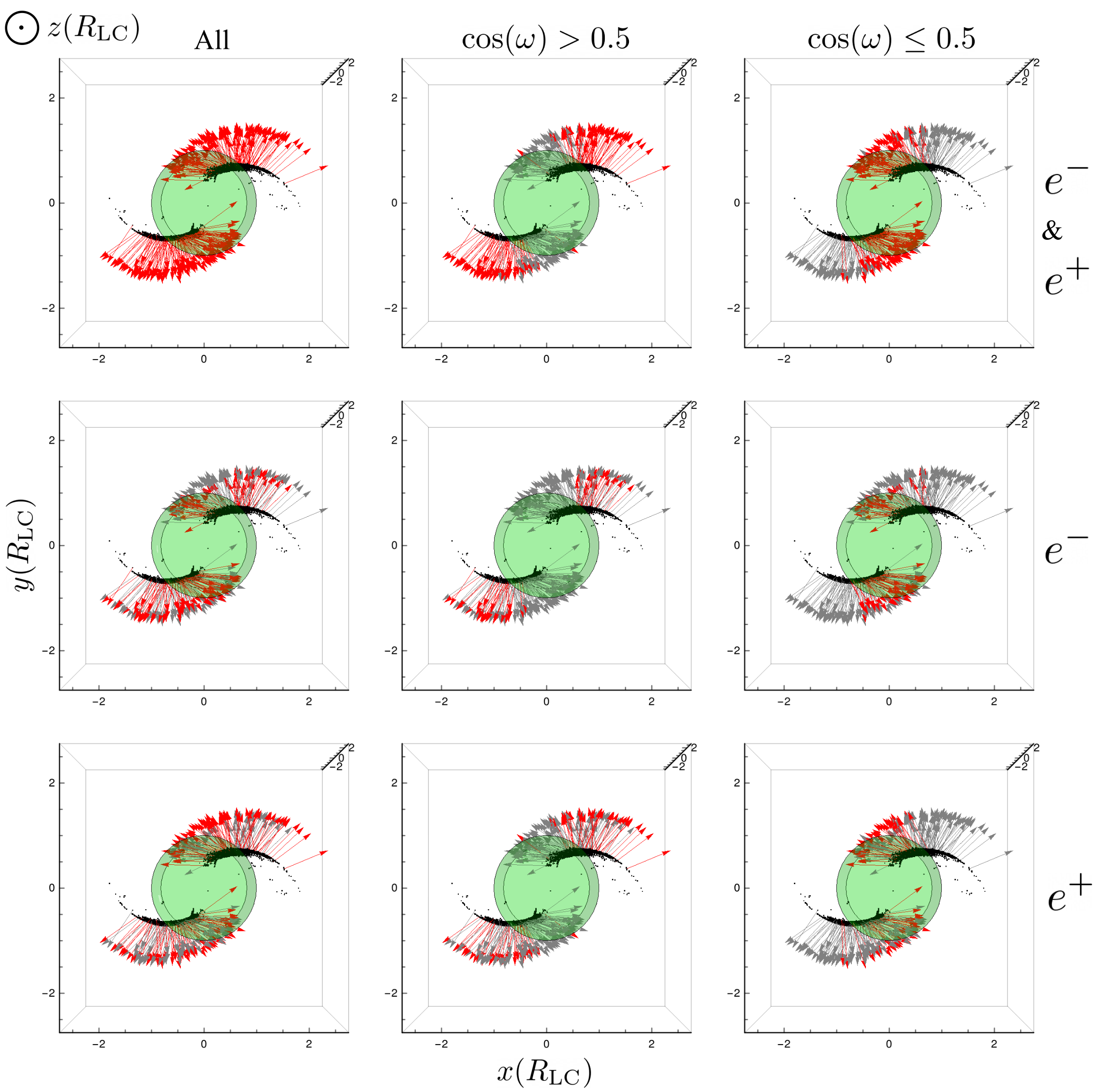}
  \end{center}
  \vspace{-0.2in}
  \caption{The particles that produce the sky-map components shown in Figure~\ref{fig:eSI_sky-maps_ele-pos_in-out} seen from the rotational axes. The black points denote the particles that produce the highest 95\% of the total emission. The red and gray arrows denote the particle velocities, i.e., photon direction. The particles with the red arrows in each panel denote the particles that produce the sky-map components shown in Figure~\ref{fig:eSI_sky-maps_ele-pos_in-out}. The green cylinder denotes the LC.}
  \label{fig:eSI_3d-from-above_ele-pos_in-out}
  \vspace{0.0in}
\end{figure*}

\begin{figure}[!tbh]
\vspace{0.0in}
  \begin{center}
    \includegraphics[width=1.0\linewidth]{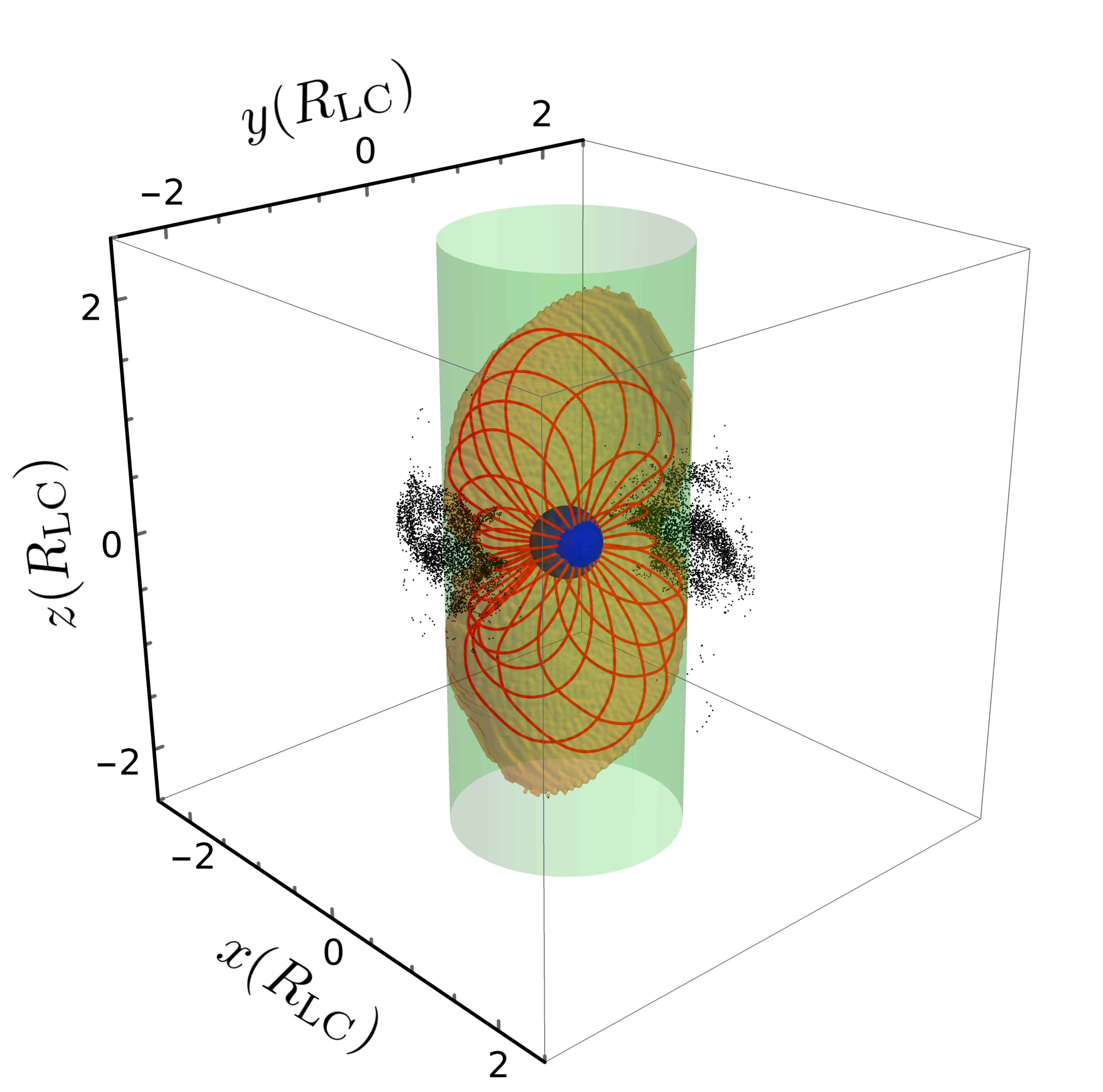}
  \end{center}
  \vspace{-0.2in}
  \caption{The separatrix (transparent orange surface), the last close field lines (red lines), and the particles that produce the highest 95\% of the total $\gamma$-ray emission for the \textit{extended separatrix zone} model of $\alpha=75^{\circ}$ and $\ed\approx 10^{38}\rm \, erg\; s^{-1}$. Reconnection and dissipation starts inside the LC (the LC delineated by the green cylinder) near the rotational equator.}
  \label{fig:separatrix3d_a75_eSZ}
  \vspace{0.0in}
\end{figure}

\begin{figure*}[!tbh]
\vspace{0.0in}
  \begin{center}
    \includegraphics[width=0.8\linewidth]{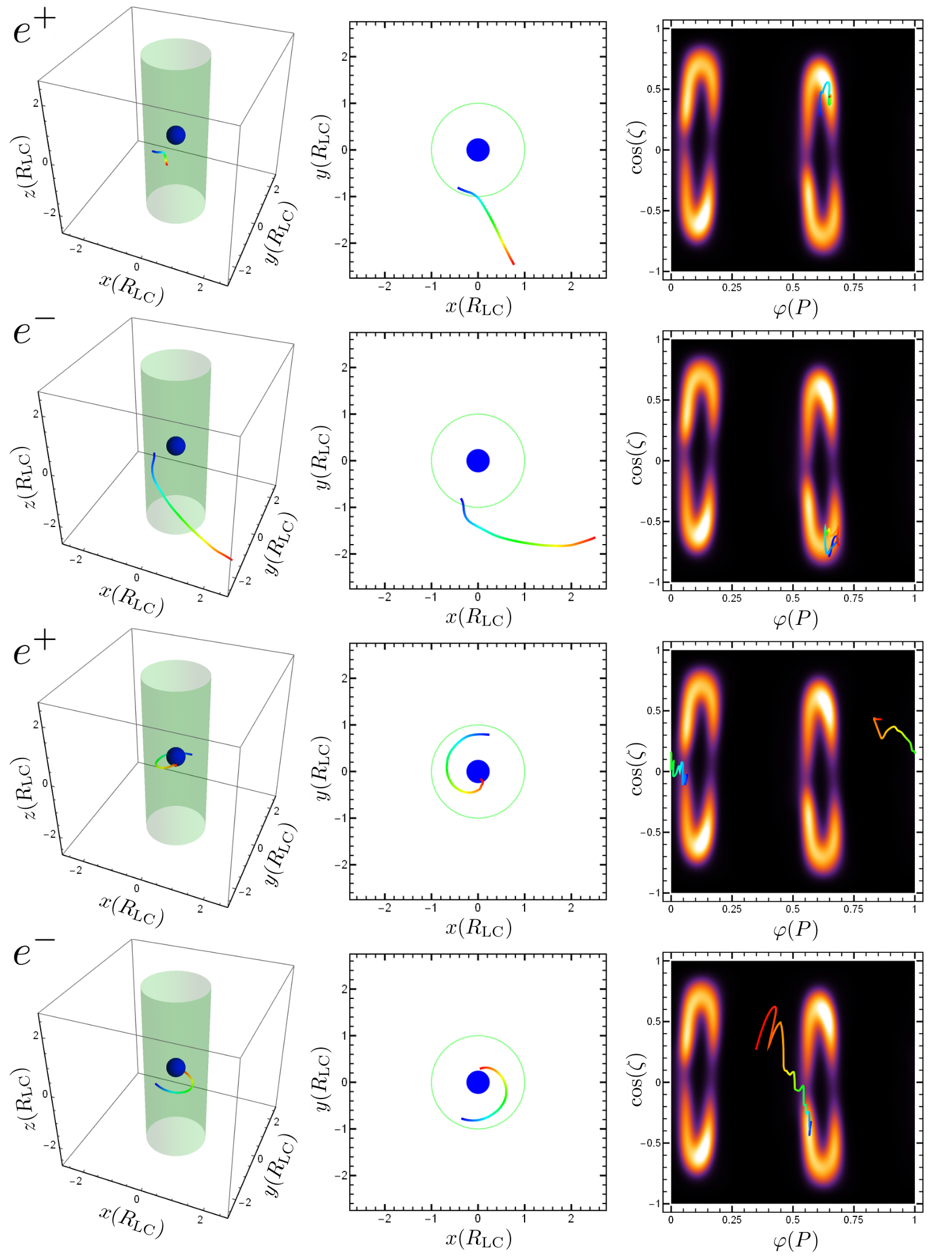}
  \end{center}
  \vspace{-0.2in}
  \caption{Representative particle orbits that contribute to different sky-map components for the \textit{extended separatrix zone} model of $\alpha=75^{\circ}$ and $\ed\approx 10^{38}\rm erg\; s^{-1}$. The orbits are plotted in the 3D space (left-hand column), on the projected $x-y$ plane (middle column), while their traces on the corresponding sky map are shown in the right-hand column. The color progression along the orbits/traces denotes the time arrow (from blue to red). The green cylinder/circle in the left-hand and middle columns denote the LC.}
  \label{fig:orbits_eSZ}
  \vspace{0.0in}
\end{figure*}

\begin{figure*}[!tbh]
\vspace{0.0in}
  \begin{center}
    \includegraphics[width=1.0\linewidth]{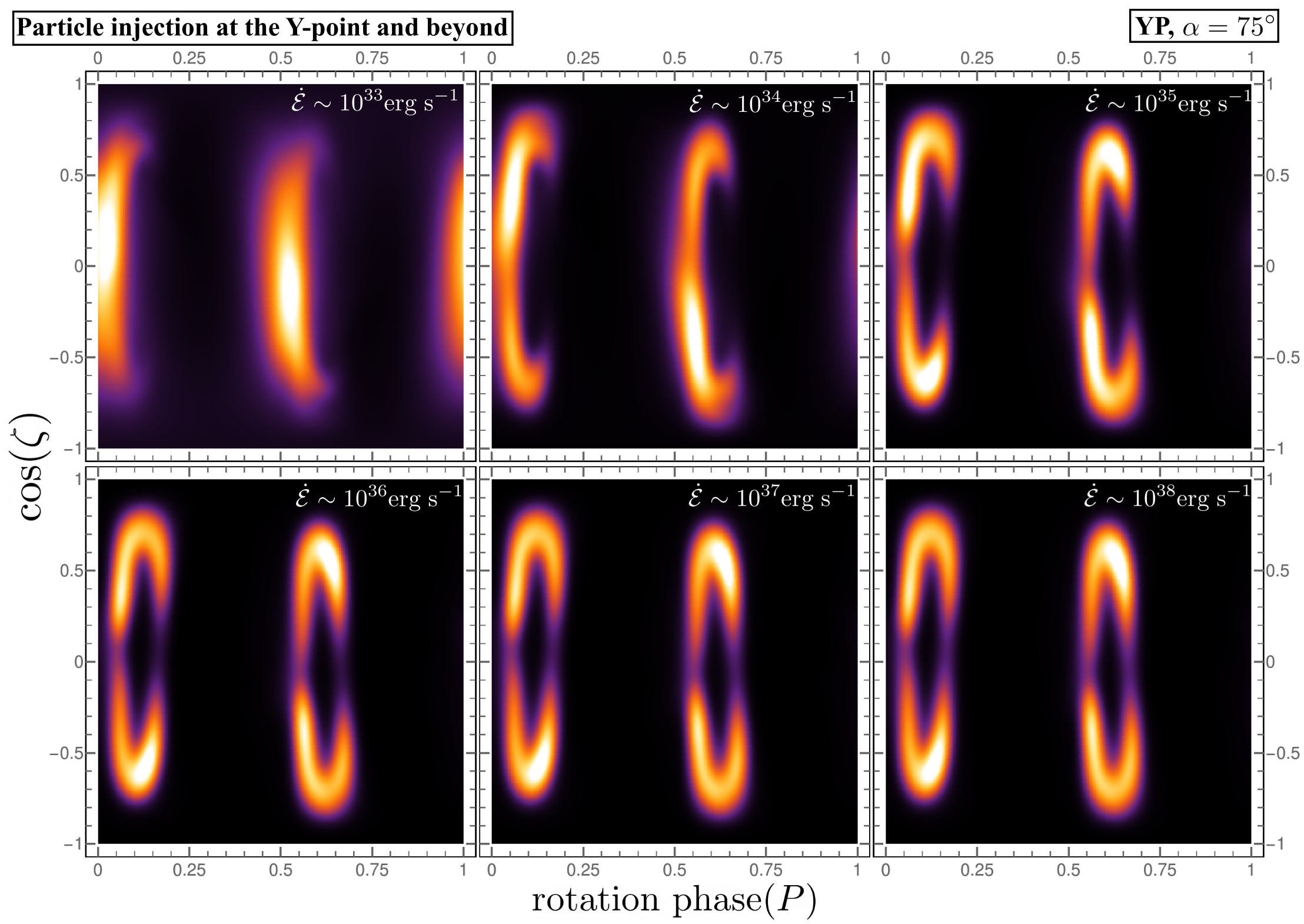}
  \end{center}
  \vspace{-0.2in}
  \caption{The sky maps for the \textit{extended separatrix zone} model of $\alpha=75^{\circ}$ and for the indicated $\ed$ values.}
  \label{fig:eSI_sky-maps_edots}
  \vspace{0.0in}
\end{figure*}

\begin{figure*}[!tbh]
\vspace{0.0in}
  \begin{center}
    \includegraphics[width=1.0\linewidth]{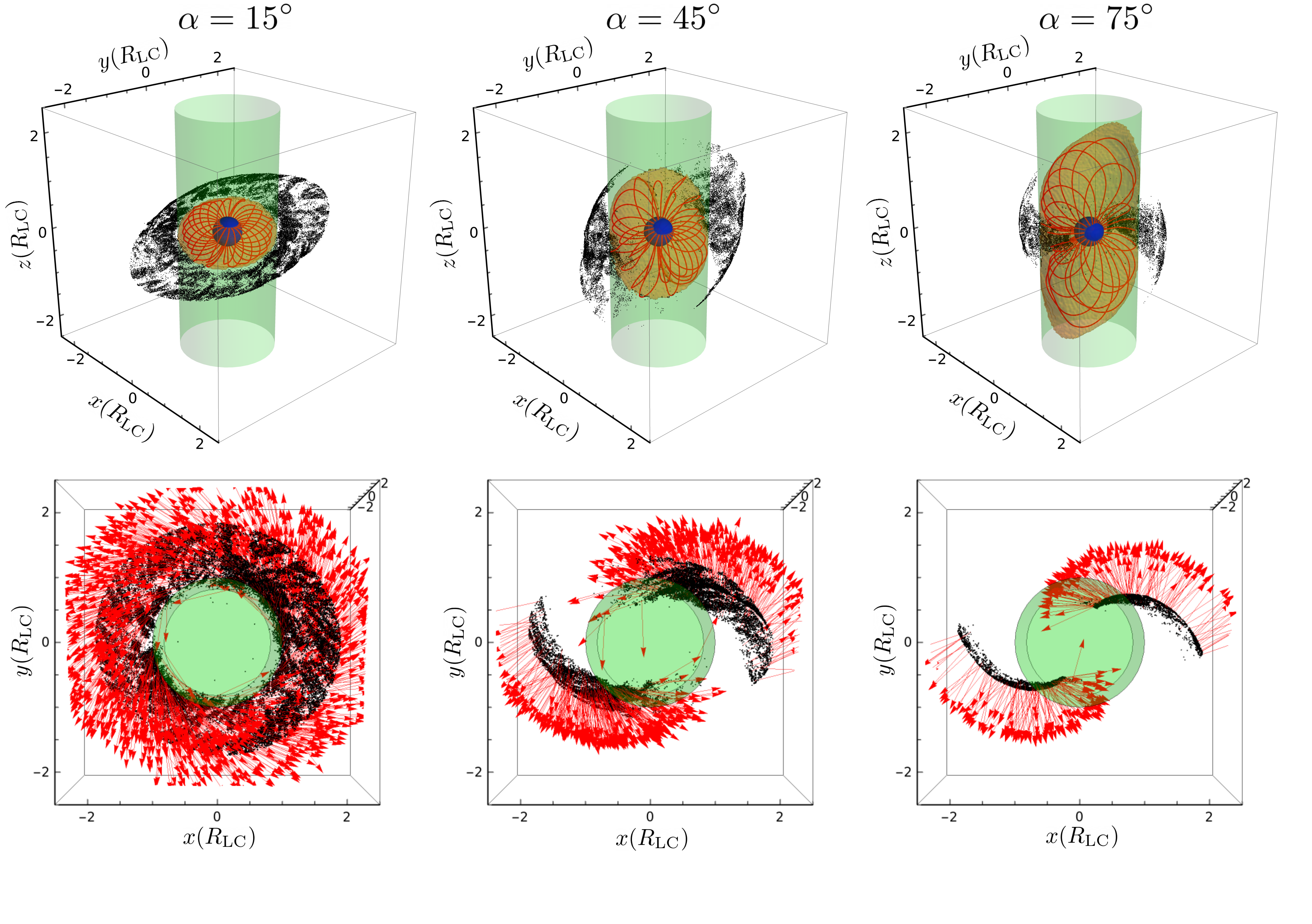}
  \end{center}
  \vspace{-0.2in}
  \caption{\textbf{Top row:} The separatrix (transparent orange surfaces), the last close field lines (red lines), and the particles that produce the highest 95\% of the total $\gamma$-ray emission (black points) for the \textit{separatrix zone} model number 6, $\ed\approx 10^{38}\rm erg\; s^{-1}$, and for the indicated $\alpha$ values. \textbf{Bottom row:} The same model cases as those presented in the top row but from the point of view that is located along the rotational axis. The red arrows denote the particle velocities, i.e., photon direction. For $\alpha=15^{\circ}$, no high emitting particles appear inside the LC. For $\alpha=45^{\circ}$, high emitting particles start appearing inside the LC near the rotational equator, which implies that the Y point in that region is formed inside the LC. This effect increases gradually with $\alpha$. Thus, for $\alpha=75^{\circ}$ the Y point, i.e., the origin of reconnection, near the rotational equator, is located well inside the LC, at $\approx 0.8 R_{\rm LC}$. High-emitting particles inside the LC introduce additional features in the sky maps, e.g., ring-type structures, that result in secondary peaks of the $\gamma$-ray light curves. Nonetheless, for the \textit{separatrix zone} model, the relative number of particles outside the LC is higher compared to the \textit{extended separatrix zone} model (see Figures~\ref{fig:separatrix3d_a75_eSZ} and \ref{fig:eSI_3d-from-above_ele-pos_in-out}), which suppresses the corresponding four-peaked pulse profiles.}
  \label{fig:SZ_separaratrix_topviews}
  \vspace{0.0in}
\end{figure*}

\subsection{To the Y-point and beyond}\label{sec:ecs}

\subsubsection{The extended separatrix zone model}\label{sec:eSZ}

The importance of the \textit{separatrix zone} model and the variation of the corresponding
pair-production efficiency with $\ed$ is proposed because this particle
population is the main contributor of the particles that enter the ECS zone
and regulate the corresponding energy dissipation and
$\gamma$-ray emission. However, this regulation can, in principle, be achieved by considering a varying (with $\ed$) particle injection that takes place directly in the dissipative region, i.e., regions of high $\eacc$ values (instead of a varying particle injection that takes place in the separatrix zone). Typically, these regions are located at the
tip of the closed zone, i.e., the so-called Y-point region and ECS. Local
studies in the ECS have indicated that this mechanism might operate
\citep{2019ApJ...877...53H} in extreme pulsars such as the Crab. The Y-point and ECS regions are the extensions
of the separatrix zone, which simply means that the difference between the
two approaches is that in the \textit{separatrix zone} model, the regulation is achieved mainly by the
variation of the pairs that are produced inside the Y-point, i.e., closer to
the star, while in the \textit{extended separatrix zone} model by the variation of the pairs that are
produced at and beyond the Y-point. i.e., closer to the LC.

In order to explore the response of the extended separatrix zone particle injection, we produce
models that have two populations of injected particles. The first population,
\textbf{P1}, is uniformly\footnote{Uniformly here means that the only
criterion is the local magnetization \citepalias{2018ApJ...857...44K}.} injected
along all the magnetic field lines up to $r=0.7R_{\rm LC}$. The second
population, \textbf{P2}, is injected for $0.5R_{\rm LC}<r<2R_{\rm LC}$
wherever $\eacc>0.01B_{\rm LC}$ and $\eacc/E>0.1$\footnote{{As we discuss below, there are cases where the Y point is formed inside the LC, and therefore, our scheme allows the P2 population to be injected even within the LC in places where the accelerating electric field components are significant, i.e., in the ECS. Enforcing the particle injection strictly outside the LC when the ECS is formed inside the LC is not physically justified.}}, where $E$ is the total local electric field. The total particle
injection rates of P1, $\ir_1$, are kept fixed at values, which alone can
sustain the FF field solution, while the total particle injection rates of
P2, $\ir_2$, freely varies. In simple terms, the role
of P1 in the \textit{extended separatrix zone} model is equivalent to the role of the particle populations
injected in the open zone and closed zone in the \textit{separatrix zone} model, while P2, in the
\textit{extended separatrix zone} model acts similarly to the particle population injected in the separatrix zone in the \textit{separatrix zone} model.

\subsubsection{Sky Maps and Pulse Profile Morphology}\label{sec:eSZ_skymaps}

The \textit{extended separatrix zone} models corresponding to low $\ir_2$ values provide results that are
very similar to the \textit{separatrix zone} ones with low $\ir_{\rm SZ}$ values. Nonetheless, for
high $\ir_2$ values, the higher $\alpha$ models start producing sky-map, i.e., $\gamma$-ray light-curve, patterns where the four-peak feature discussed above (see Figure~\ref{fig:skymap75}) appears enhanced.

In the top-left panel of Figure~\ref{fig:eSI_sky-maps_ele-pos_in-out}, we plot the total sky map corresponding to a YP parameter set, i.e., $\ed~\approx~10^{37}\rm \, erg\;s^{-1}$ for an \textit{extended separatrix zone} model
with $\alpha=75^\circ$, $\ir_1=20F_{\rm GJ}^0$, and
$\ir_2=7F_{\rm GJ}^0$. These values produce a model with a FF-ness similar to \textit{separatrix zone} model No.~6, i.e., the distribution of the encountered $V$ reaches similar levels. The $V$ distribution is plotted in the top panel of Figure~\ref{fig:voltage_distribution} with a magenta-colored line.

The rest of the panels of Figure~\ref{fig:eSI_sky-maps_ele-pos_in-out} show sky-map components corresponding to the indicated particle populations (see below and Figure caption). Similarly, in the top-left panel of Figure~\ref{fig:eSI_3d-from-above_ele-pos_in-out}, we plot in 3D space (viewed from the rotational axis) the particles, i.e., black points, that produce the highest 90\% of the total $\gamma$-ray luminosity and are responsible for the top-left sky-map signal of Figure~\ref{fig:eSI_sky-maps_ele-pos_in-out}. The red arrows indicate the velocity directions, i.e., the direction of the emitted $\gamma$-ray photons, corresponding to a random subset of particles. There is one-to-one correspondence between the panels of Figure~\ref{fig:eSI_sky-maps_ele-pos_in-out} and Figure~\ref{fig:eSI_3d-from-above_ele-pos_in-out}. Therefore, the sky-map components plotted in Figure~\ref{fig:eSI_sky-maps_ele-pos_in-out} are produced by the particles with velocity directions denoted by the red arrows in Figure~\ref{fig:eSI_3d-from-above_ele-pos_in-out}. Figures~\ref{fig:eSI_sky-maps_ele-pos_in-out} and \ref{fig:eSI_3d-from-above_ele-pos_in-out} indicate that the left (right)-hand sides of the ring sky-map patterns are mainly produced by the particles that lie inside (outside) the LC. Nonetheless, a more careful consideration shows that the particles with $\cos(\omega)\lesssim 0.5$ $(\cos(\omega)\gtrsim 0.5)$ produce the left (right)-hand sides of the rings, where $\omega$ is the angle between the particle velocity, i.e., photon direction, and the radial direction at the corresponding particle position. Furthermore, Figures~\ref{fig:eSI_sky-maps_ele-pos_in-out} and \ref{fig:eSI_3d-from-above_ele-pos_in-out} show the sky-map contributions of $e^{-}$ and $e^{+}$. Thus, $e^{-}$ and $e^{+}$ are responsible for different segments of the ring-type patterns. \citet{2016MNRAS.457.2401C} had also reported different $e^-$ and $e^+$ light-curve components.

Figure~\ref{fig:eSI_3d-from-above_ele-pos_in-out} indicates that a significant high-energy emitting particle population is located well inside the LC. This happens because, in this case, the so-called Y point near the rotational equator is formed inside the LC. This implies that the formation of the ECS and, therefore, reconnection starts inside the LC. In Figure~\ref{fig:separatrix3d_a75_eSZ}, we plot, in the 3-dimensional space, the separatrix, i.e., orange transparent surface, together with the high-energy emitting particles, i.e., black points. The red lines denote the last closed field lines, which lie on the separatrix. Figure~\ref{fig:separatrix3d_a75_eSZ} shows that the separatrix around the rotational equator does not touch the LC, which vividly demonstrates the opening of the magnetic field lines and, therefore, the origin of the ECS inside the LC. Figure~\ref{fig:eSI_3d-from-above_ele-pos_in-out} shows that the particles lying inside the LC have significant toroidal velocity components. The onset of the ECS inside the LC forces the ECS to cross the LC. This LC crossing considerably changes the velocity flow of the high-energy particles adding complexity to the sky-map features.

In Figure~\ref{fig:orbits_eSZ}, we plot characteristic types of orbits and their contribution to the sky maps. More specifically, the left-hand column shows the 3D trajectories, while the middle column shows the trajectory projections on the $x-y$ plane. The right-hand column shows the corresponding sky map and the corresponding trajectory trails. The color along the trajectories/trails denotes the time arrow, i.e., from blue to red. The particles inside the LC, i.e., the bottom two rows of Figure~\ref{fig:orbits_eSZ}, contribute to the left-hand side of the ring-type pattern while having high energies. However, after some time, these particles detach from the ECS, and so they stop being accelerated, and their energy decreases, considerably depressing their high-energy emission.

The magnetic field values increase as the distance from the star decreases. Therefore, the $\eacc$ values in the ECS are stronger inside the LC. As long as RRL dictates the particle energies, the relative strength of the emission between the particles inside and outside the LC, i.e., the sky-map pattern, remains unaffected by the variation of $\ed$. Nonetheless, when the RRL ceases to hold sway over the particle energies, the particles that encounter higher $\eacc$ (inside the LC) accelerate more efficiently. In Figure~\ref{fig:eSI_sky-maps_edots}, we plot, for the same \textit{extended separatrix zone} model, the sky maps corresponding to the indicated $\ed$ values. We see that for $\ed\gtrsim 10^{36}\rm erg\;s^{-1}$, where the RRL regime is dominant, the sky maps are very similar to each other. However, for lower $\ed$ values, the RRL regime gradually stops governing the particle energies (see Figure~\ref{fig:RRLR}), and therefore, the contribution of the particles that lie inside the LC becomes dominant. This results in sky-map patterns of two peaks, which, however, have very small radio-lag, $\delta$, values. Nonetheless, we note that the \textit{extended separatrix zone} model presented in Figure~\ref{fig:eSI_sky-maps_edots} is not compatible for the low $\ed$ values because the corresponding spectral cutoff energies are considerably smaller than those observed, i.e., outside the Fermi pulsar band. In any case, we note that the \textit{extended separatrix zone} model for all the $\ed$ values is still consistent with the model FP, i.e., Eq.~\eqref{eq:fp_all_pic}.

\subsubsection{The locus of the Y point region}\label{sec:eSZ_locus_of_Y-point_region}

As mentioned above, the four-peaked sky-map morphology, though weaker, is present in the corresponding \textit{separatrix zone} model. A more careful examination shows that in the highest FF-ness \textit{separatrix zone} model, i.e., model No.~6, the Y point near the rotational equator is also located well inside the LC. The panels in the top row of Figure~\ref{fig:SZ_separaratrix_topviews} are similar to Figure~\ref{fig:separatrix3d_a75_eSZ} while those in the bottom row are similar to the top-left panel of Figure~\ref{fig:eSI_3d-from-above_ele-pos_in-out} but for the \textit{separatrix zone} models number 6. We see that for $\alpha=75^{\circ}$, the separatrix surface touches the LC even in the rotational equator, which implies that all the magnetic field lines  close inside the LC. This is due to reconnecting magnetic field lines inside the LC since the Y point near the rotational equator is again located inside the LC. Nonetheless, in this case, there are relatively more particles outside the LC than those inside the LC, which makes  the left-hand side components of the ring-type features relatively less intense on the sky map. For the $\alpha=15^{\circ}$ case, the separatrix always touches the LC. In this case, the Y point line is located at the LC, so the reconnection, i.e., dissipation, starts beyond the Y point at the ECS. For $\alpha=45^{\circ}$, high-energy particles start appearing inside the LC near the rotational equator, indicating the migration of the Y point of this area inside the LC. Nonetheless, this migration is small, and the number of particles that emit inside the LC is considerably smaller than those that are located outside the LC. We note that for the \textit{separatrix zone} models, the Y-point area located inside the LC  moves outwards gradually as the FF-ness decreases, i.e., the model number decreases.

A Y point well inside the LC, except for the aligned case, i.e., $\alpha=0^{\circ}$, has been shown in several studies \citep[e.g.,][]{2014ApJ...795L..22C,2015MNRAS.448..606C,2020A&A...635A.138G,2022ApJ...939...42H,2022arXiv220902121H} in the past. More specifically, \citet{2022ApJ...939...42H}, who treated only the aligned case, reported strong azimuthal motion of high-energy particles beyond the Y point inside the LC. The results of \citet{2022arXiv220902121H} also implied that the Y point migrates gradually towards the star, inside the LC, as the particle injection rate increases. As discussed above, we do not see this effect for low $\alpha$ values, at least for the parameter values we have explored. We are also unaware of similar reports regarding the migration of the Y point inside the LC for oblique rotators, especially for high $\alpha$ values. However, we note that typically figures in the literature show the field structure for oblique rotators on the poloidal $\pmb{\mu}-\pmb{\Omega}$ plane, where the feature reported in this study is not shown.

Despite the similarities between our results for oblique-only rotators and those of other studies for aligned rotators, it remains unclear if they both describe the same physical effect. Moreover, our results and the results of \citet{2022arXiv220902121H} indicate that the Y point migrates inwards as the particle injection rate increases. This raises concerns about whether the Y point migration is due to particle inertia, which non-realistically decreases magnetization leading to the opening of further magnetic field lines.

In order to falsify this conjecture, we ran two simulations with the same particle injection scheme, but one with 
5 times higher magnetic field and the other one with 
5 times less. The magnetization $\sigma_{\rm M}=B^2/(4\pi n\Gamma m_{\rm e} c^2)$, where $\Gamma$ the Lorentz factor of the local bulk plasma flow and $n$ the local total particle, i.e., $e^{-}$ and $e^{+}$, number density, is proportional to $B/\Gamma$ since for the same prescription scheme, $n\propto B$. The ratio of the $\sigma_{\rm M}$ values at the LC near the rotational equator corresponding to the aforementioned simulations is 5.5, i.e., $\simeq 450$ and $\simeq 80$, respectively. Despite this magnetization difference, the location of the Y points and the origin of the formation of the ECS remains unaffected. Moreover, the high magnetic field model has higher magnetization than other \textit{separatrix zone} models, e.g., model No. 4, in which, however, the corresponding Y points are located at larger distances much closer to the LC than the Y point of the simulation with the higher magnetization. These findings provide strong evidence that the migration of the Y point inside the LC near the rotational equator is real and not artificially triggered by particle inertia.

Starting from the retarded vacuum field solution, an increasing particle injection, i.e., pair production, gradually leads to the development of field configurations close to the FF ones where the Y-point regions are located near the LC. Far from the vacuum solutions, the main dissipative region is located in the ECS, forming at and beyond the corresponding Y-point region. However, the gradually increasing particle injection rate starts suppressing the dissipation, i.e., the rate of reconnecting magnetic field. The region near the Y point and especially near the rotational equator of oblique rotators are the most resilient dissipative locations, which seem to react to the attempted suppression of the reconnecting accelerating electric field components by pushing the Y point inwards where, because of the higher magnetic field values there, more particles are needed to achieve the same level of suppression of dissipation, i.e., reconnection. This interpretation implies that the Y point can be pushed arbitrarily close to the stellar surface depending on the number of particles that are either injected, i.e., produced at or reach the corresponding dissipative region beyond the LC. This view appears consistent with the results presented in this study and probably other studies, but it remains a conjecture that must be further studied. This view is also consistent with the results of macroscopic FF simulations. More specifically, the utilization of numerical schemes that meticulously suppress magnetic reconnection in macroscopic FF simulations leads to solutions that have, even temporarily, the Y-points inside the LC \citep{2006ApJ...648L..51S,2009A&A...496..495K}. Furthermore, in macroscopic magnetosphere models, the Poynting flux, i.e., $\ed$, for high $\alpha$ values starts deviating from the relation $\ed=3/2 \ed_{V_{90}}(1+\sin^2 \alpha)$\footnote{It is noted that $\ed_{V_{90}}$ denotes the spin-down power for the perpendicular vacuum retarded dipole solution.} \citep{2006ApJ...648L..51S}, toward higher values \citep[e.g.,][]{2013MNRAS.435L...1T}, possibly the result of the opening of additional magnetic flux. A more detailed study of this effect, i.e., the migration of the Y point well inside the LC, is beyond the scope of this paper and will be addressed in a future study. Nonetheless, we note that our investigation not only indicates that the migration of the Y point inwards is a physical and not an artificial effect but also reveals the corresponding observational consequences.

\section{Summary and Discussion}\label{sec:conclusions}

Fermi-LAT provides the patterns of the $\gamma$-ray light curves for hundreds of pulsars, setting strong constraints on the field geometries and the location of the emitting magnetosphere regions. Yet, Fermi-LAT also provides spectral information, i.e., $L_{\gamma},~\ec$, setting additional constraints on the $\gamma$-ray efficiency and the accelerating electric field components, $\eacc$ in the dissipative regions. Recently, using data from \citet{2013ApJS..208...17A} and later from \citet{2022arXiv220111184F}, we showed that the Fermi $\gamma$-ray pulsars lie on a FP that relates $\ed,~L_{\gamma},~B_{\star},~\ec$, consistent with curvature radiation \citep{2019ApJ...883L...4K,2022ApJ...934...65K}. This FP implies that the four observables are not independent but are related by a relation that describes a 3D plane (the fundamental plane of $\gamma$-ray pulsars) embedded in 4D.

On the other hand, previous studies of global macroscopic and PIC models indicated that the main features of the Fermi $\gamma$-ray light curves are consistent with magnetosphere structures close to the FF ones and emission that originates near the ECS beyond the LC. Furthermore, the plasma conductivity in the dissipative region, i.e., ECS, which is associated with the local particle number density, is the regulating factor of the spectral properties, i.e., $L_{\gamma},~\ec$ \citep{2014ApJ...793...97K,2015ApJ...804...84B,2017ApJ...842...80K,2018ApJ...857...44K,2018ApJ...855...94P}.

In the present study, we developed a series of innovative PIC models of pulsar magnetospheres that reproduce a broad spectrum of the observed phenomenology. Our models were built considering the aforementioned remarks. More specifically, the requirement of the FF field configuration implies that the global particle injection rate should be sufficiently high to support the development of such FF field configurations. However, regulating the spectral properties requires a variable particle loading of the dissipative region, i.e., the ECS beyond the Y-point locus. Our studies had indicated that the particles that reach the ECS region are principally produced in the broader separatrix zone that separates the open and closed magnetic field lines \citepalias{2018ApJ...857...44K}.

Thus, we developed models in which the particle injection rate along most open and closed field lines is fixed to some value of the order of $\simeq 10F_{\rm GJ}^0$, which ensures the development of a global FF field configuration. On the other hand, we varied the particle injection rate along the magnetic field lines around a narrow zone around the separatrix surface. In these models, the particle injection takes place from the stellar surface up to a distance $0.7R_{\rm LC}$.

The higher the particle injection rate in the separatrix zone is and the thinner this zone is, the lower the voltage the particles encounter, i.e., the higher FF-ness is. We ran models for three inclination angle values, i.e., $\alpha=15^{\circ},~45^{\circ},~75^{\circ}$ considering twelve sets of $B_{\star},~P$ values that delineate the parameter-space area of the observed Fermi YPs and MPs. We note that one $B_{\star},~P$ set for MPs and one set for YPs correspond to $\ed$ values slightly below the lowest values Fermi has detected.

For each of these models, we derived the corresponding $\gamma$-ray emission patterns, i.e., sky maps, and from these, the $\gamma$-ray light curves corresponding to different observers. Assuming that the radio emission of YPs originates proximate to magnetic poles near the stellar surface, we derived $\delta,~\Delta$ values for a suite of models, which reproduce the observed $\delta-\Delta$ correlation reported in \citet{2013ApJS..208...17A} with remarkable fidelity. Moreover, we calculated the model spectra and the corresponding $L_{\gamma},~\ec$ values. This led to the derivation of the model FP, which was in agreement with the theoretical FP for curvature radiation.

Moreover, the model FP and the observed one reported in  \citet{2022ApJ...934...65K} are near parallel to each other, i.e., the dependencies between the various variables are similar in the two cases. However, the observed FP indicates $L_{\gamma}$ values higher than those corresponding to the model FP by a factor $\simeq 3.5$. However, our models indicate that the $f_{\rm b}$ values are usually smaller than 1, especially along the line of sights, i.e., $\zeta$ values, where the emissivity is stronger. The adopted $f_{\rm b}$ value in \citet{2013ApJS..208...17A} and \citet{2022arXiv220111184F} was 1, which implies that according to our models, the $L_{\gamma}$ values reported in the Fermi data are overestimations of the actual values. Since we do not know the $\alpha$ and $\zeta$ of Fermi pulsars, we cannot add the $f_b$ correction to the data.  This explains, at least partially, the apparent inconsistency between the model FP and the observed one. Additionally, recent studies indicate that the moment of inertia, $I_{\rm M}$ especially of MPs, is slightly higher than $10^{45} \rm g\;cm^2$, which is the adopted value in the Fermi results in \citep{2013ApJS..208...17A,2022arXiv220111184F}. This implies that the reported $\ed,~B_{\star}$ values are lower than the actual ones bringing the observed FP even closer to the model one. Finally, we note that the model $L_{\gamma}$ values have been calculated considering the emission up to 2$R_{\rm LC}$, which does not include possible subdominant emissivity at larger radii. Even though this emission component is expected to be small, it ought to account for some of the normalization offset between the model and observed FPs.

In \citet{2019ApJ...883L...4K}, we claimed that the RRL regime is a necessary condition for the FP theory. We relaxed this condition in \citet{2022ApJ...934...65K}, showing that the RRL regime is sufficient but not necessary. Indeed, our simulations show that emission at the RRL regime occurs for $\ed\gtrsim 10^{34}\rm \, erg\; s^{-1}$ for YPs and $\ed\gtrsim  10^{33}\rm \, erg\; s^{-1}$ for MPs. Pulsars remain on the FP independent of whether the radiating particles are in the RRL regime or not. Nonetheless, the RRL regime affects the pulsar position on the FP. We will discuss this effect and show the observational consequences in a forthcoming paper.

In addition, we explored solutions in which an enhanced particle injection takes place in the outer magnetosphere directly in high-dissipation regions, i.e., the Y point and the ECS emanating from it. These simulations revealed that the increasing particle injection, especially for high $\alpha$ values, leads to a gradual migration of the Y-point region located near the rotational equator inside the LC. The observational consequence of this effect is the formation of four distinct light-curve peaks -- which are manifestly inconsistent with Fermi-LAT observations. The additional emission components come from the particles that emit inside the LC. These particles are accelerated in the ECS region, which, in these cases, originates inside the LC. The velocity pattern changes from mainly toroidal inside the LC to gradually radial outside the LC. The effect of the Y-point migration inside the LC has been reported before, but primarily for the aligned rotator, which does not produce light-curve signals. Checking the behavior of our simulations for different magnetization values, we have concluded that this effect is not a result of particle inertia. The effect appears to become less significant but not absent when the enhanced particle injection takes place well inside the LC, i.e., in the \textit{separatrix zone} model. An important aspect that determines the intensity of this effect is the relative balance between the particles that emit inside versus those outside the LC. Producing particles well beyond the Y point in the ECS region reduces the acceleration efficiency of the area, enhancing the relative contribution of the inner emitting component, i.e., inside the LC. The regulation of the $\gamma$-ray emission and the corresponding spectra indicates varying (with $\ed$) particle multiplicities in the dissipation region.

The \textit{separatrix zone} model presented in this study demonstrates the importance of the efficiency of the pair production in the separatrix zone regarding the regulation of the emitted $\gamma$-ray emission. Nonetheless, the adopted scheme that incorporates three distinct zones is rather simplistic. In reality, a continuous distribution of the pair production efficiency near the separatrix surface is expected, with significant angular dependence. Even though our study indicated that the efficiency of the pair production in the separatrix zone and its dependence on $\ed$ is crucial, it remains unclear whether the pair-production efficiency is indeed very different in the three different zones even though we did not manage to develop solutions similar to models No.~1 using uniform particle injection. Our scheme, which assumed very different pair production efficiencies, was an idealized one intended to demonstrate the importance of charge carriers in the various zones. At the same time, it was numerically expedient because it reduced the total number of particles in regions where particle injection was inconsequential.

Our models successfully reproduce the broad properties of the observed phenomenology of Fermi pulsars, from the FP of $\gamma$-ray pulsars to gross features of the corresponding pulse profiles. Their success reveals the operational regime of the underlying (micro)-physical mechanisms compatible with the observations. However, our models remain
mainly descriptive, lacking a profound physical justification that provides a rigorous quantification that validates the trends indicated in this study.

\citet{2013MNRAS.429...20T} showed that different polar-cap current regimes, determined by the corresponding global current, $J/\rho c$, values, support different pair-production efficiencies, which immediately implies that the various zones should be along the magnetic field lines corresponding to the different current-regimes. For the aligned rotator, the current structure matches the adopted scheme in this study, but for higher $\alpha$ values, the developed current structure asymmetries bear possibly variable pair efficiencies along the separatrix zone. Moreover, \citet{2013MNRAS.429...20T} implemented one-dimensional simulations limiting the applicability of their results, especially near the separatrix conductive surface.
The proper approach requires a multi-dimensional study that is expected to provide a much more realistic treatment compared to what has been done so far, something that will allow the exploration of the interaction of the different pair-production operation regimes, including the influence of the conductive boundaries.

\noindent \emph{Acknowledgments.} We would like to thank the International Space Science Institute (ISSI) for providing financial support for the organization of the meeting of the ISSI Team that was led by I. Contopoulos and D. Kazanas. C.K., Z.W., and D.K. are supported by the
Fermi Guest Investigator program under award numbers 80NSSC21K1999, 80NSSC21K2001, and 80NSSC22K1908, the NASA Theory Program under award number 80NSSC22K1267, and the NASA Astrophysics Data Analysis Program under the award number 80NSSC23K0462. The material is based upon work supported by
NASA under award number 80GSFC21M0002.

\end{document}